%% file: paper.tex
\documentclass[useAMS,usenatbib]{mn2e}

\usepackage{graphicx}
\usepackage{psfig}
\usepackage{amssymb,amsmath}
\usepackage{mathrsfs}
\usepackage{times}
\usepackage{setspace}
\usepackage{color}

\include{journal}

\newcommand{\begit}{\begin{itemize}}
\newcommand{\enit}{\end{itemize}}
\newcommand{\begen}{\begin{enumerate}}
\newcommand{\enen}{\end{enumerate}}

\newcommand{\beq}{\begin{equation}}
\newcommand{\eeq}{\end{equation}}
\newcommand{\beqa}{\begin{eqnarray}} 
\newcommand{\eeqa}{\end{eqnarray}} 
\def\lesssim{\mathrel{\hbox{\rlap{\hbox{\lower5pt\hbox{$\sim$}}}\hbox{$<$}}}}
\def\gtrsim{\mathrel{\hbox{\rlap{\hbox{\lower5pt\hbox{$\sim$}}}\hbox{$>$}}}}

\title[Radiation Pressure Driven Shells and Clouds]
{Dynamics of Dusty Radiation Pressure Driven Shells and Clouds: Fast Outflows from 
Galaxies, Star Clusters, Massive Stars, \& AGN}

\author[Thompson et al]{
Todd A.~Thompson$^1$,
Andrew C.~Fabian$^2$, 
Eliot Quataert$^3$,
and Norman Murray$^{4,5}$\\
$^1$Department of Astronomy and Center for Cosmology \& Astro-Particle Physics,
The Ohio State University, Columbus, Ohio 43210\\ 
$^2$Institute of Astronomy, Madingley Road, Cambridge CB3 0HA\\
$^3$ Department of Astronomy and Theoretical Astrophysics Center, 
University of California, Berkeley, CA 94720-3411\\
$^4$Canadian Research Chair in Astrophysics\\
$^5$Canadian Institute for Theoretical Astrophysics, 
The University of Toronto, 60 St.~George Street, Toronto 
Ontario M5S 3H8
}

\begin{document}
\maketitle
\label{firstpage}
\begin{abstract}
It is typically assumed that radiation pressure driven winds are accelerated to an asymptotic velocity of $v_\infty\simeq v_{\rm esc}$, where $v_{\rm esc}$ is the escape velocity from the central source.  We note that this is not the case for dusty shells and clouds.   Instead, if the shell or cloud is initially optically-thick to the UV emission from the source of luminosity $L$, then there is a significant boost in $v_\infty$ that reflects the integral of the momentum absorbed as it is accelerated.  For shells reaching a generalized Eddington limit, we show that $v_\infty\simeq(4R_{\rm UV}L/M_{\rm sh} c)^{1/2}$, in both point-mass and isothermal-sphere potentials, where $R_{\rm UV}$ is the radius where the shell becomes optically-thin to UV photons, and $M_{\rm sh}$ is the mass of the shell.  The asymptotic velocity significantly exceeds $v_{\rm esc}$ for typical parameters, and  can explain the $\sim1000-2000$\,km s$^{-1}$ outflows observed from rapidly star-forming galaxies and active galactic nuclei if the surrounding halo has low gas density.  Similarly fast  outflows from massive stars can be accelerated on $\sim{\rm few}-10^3$\,yr timescales.  These results carry over to clouds that subtend only a small fraction of the solid angle from the source of radiation and that expand as a consequence of their internal sound speed.  We further consider the dynamics of shells that sweep up a dense circumstellar or circumgalactic medium.  We calculate the ``momentum ratio" $\dot{M} v/(L/c)$ in the shell limit and show that  it can only significantly exceed $\sim2$ if the effective optical depth of the shell  to re-radiated FIR photons is much larger than unity. We discuss simple prescriptions for the properties of galactic outflows  for use in large-scale cosmological simulations.  We also  briefly discuss applications to the dusty  ejection episodes of massive stars, the disruption of giant molecular clouds, and  AGN.  
\end{abstract}

\section{Introduction}
\label{section:introduction}
In the galactic context, radiation pressure on dust grains has been discussed 
as a mechanism for launching galactic-scale winds in starbursts and rapidly star-forming 
galaxies \citep{harwit,chiao72,ferrara90,mqt,mmt,hopkins12_wind, krumholz_thompson13,davis_kt}, in disrupting the 
dusty gas in individual star clusters \citep{harwit,odell,scoville01,krumholz_matzner,mqt10}, in 
launching fast outflows from AGN \citep{scoville_norman, roth}, 
in setting the  $M-\sigma$ relation \citep{fabian99,mqt}, and in supporting 
starbursts and AGN disks against their own self-gravity 
\citep{ferrara93,scoville,tqm,andrews_thompson,krumholz_thompson12}. 
Radiation pressure and momentum injection 
by supernovae and stellar winds plays an important role in models of feedback in 
star-forming galaxies 
\citep{tqm,hopkins11_feedback,ostriker_shetty,hopkins12_ism,faucher-giguere_feedback}.

In the stellar context, dusty shells are produced during the 
eruptions of supernova impostors and luminous blue variables, including $\eta$-Carinae \citep{davidson_humphreys,smith_gehrz,smith_etal_etacar,smith_etacar_LH,smith_etacar}, 
ultra-bright supernovae such as SN 2006gy \citep{miller_echo}, and
SN 2008S-like transients \citep{kochanek_dustformation, kochanek_impostor12,prieto_2008S,thompson_2008S,prieto_dust,bond}.
Dusty shell formation and dynamics are also important to the phenomenology of  
R Coronae Borealis stars (e.g., \citealt{gillett_rcorbor}), and 
continuous dusty winds are also produced generically by
AGB stars \citep{ivezic_elitzur95, dusty} and OH-IR stars
and cool hypergiants like IRC+10420 (e.g., \citealt{ridgway,humphreys97}) .  

The dynamics of radiation pressure-driven shells and clouds has been treated by
a number of authors. Here, we provide a brief discussion that makes clearer
the asymptotic velocity and momentum of an initially optically-thick shell or cloud
and connect with observations in several contexts, but with a focus on rapidly star-forming galaxies.
In particular, we critically examine the assumption that the asymptotic velocity 
of a radiation pressure driven shell or cloud is of order the escape velocity from the central body.
This expectation follows from consideration of the momentum equation for a continuous
time-steady radiation pressure driven flow with constant opacity from a point mass 
$M$ and luminosity $L$ (e.g., eq.~9 of \citealt{salpeter_radwind}): 
\beq
v \frac{dv}{dr}=-\frac{GM}{r^2}+\frac{\kappa L}{4\pi r^2 c} 
\Longrightarrow v_\infty^2=v_{\rm esc}^2(R_0)\left(\Gamma-1\right),
\label{cont_simple}
\eeq
where $R_0$ is the initial radius, $\Gamma=L/L_{\rm Edd}=L/(4\pi GMc/\kappa)$, and where 
the initial velocity of the medium has been neglected.  For line-driven winds from hot sources
(e.g., main sequence O stars, Wolf-Rayet stars, the central sources of planetary nebulae, AGN) the opacity is dominated 
by a forest of Doppler-shifted metal lines, $R_0$ is typically of order the radius of the 
illuminating object, and the effective opacity is usually a multiplicative factor times $\kappa_T$,
yielding the observed correlation between $v_{\rm esc}$ and $v_\infty$ for some 
object classes (e.g., hot star winds; \citealt{abbott78}).  For continuous winds from AGB stars, $R_0$
corresponds to the dust sublimation/formation radius ($R_{\rm sub}$; 
here the flow becomes super-Eddington)
and $\kappa$ is the flux-mean dust opacity as a function of radius \citep{ivezic_elitzur95, dusty}.  
For dusty flows that are optically-thick to the emission from the central star, most of the momentum
is absorbed in a narrow layer near $R_{\rm sub}$, wind material at larger 
radii is shielded from the central source, and $v_\infty\simeq v_{\rm esc}(R_{\rm sub})\Gamma(R_{\rm sub})^{1/2}$.

As we discuss in 
more detail below, the dynamics of a single geometrically thin dusty shell or cloud is different because as it
expands it goes through an extended phase where it is optically-thick to the assumed
incoming UV photons from the source, but optically-thin to the re-radiated 
IR emission from the grains.  For a shell that subtends $4\pi$ 
in this so-called single scattering limit (see eq.~\ref{vx}, below),
\beq
M_{\rm sh} v \frac{dv}{dr}=-\frac{GMM_{\rm sh}}{r^2}+\frac{L}{c} 
\Longrightarrow v_\infty^2\sim\frac{R_{\rm UV}L}{M_{\rm sh}c},
\label{shell_simple}
\eeq
where $M_{\rm sh}$ is the mass of the shell, $R_{\rm UV}$ is the radius at which the shell 
becomes optically-thin to the UV radiation, and we have assumed $R_{\rm UV}\gg R_0$ and that 
$L/c\gg GMM_{\rm sh}/R_0^2$.   The lack of radial dependence to the radiation pressure driving
term shifts the momentum deposition to large scales, $\sim R_{\rm UV}$, instead of $R_0$ as in 
equation (\ref{cont_simple}).  Because the shell sees the entire source luminosity $L$ during its
entire evolution, it reaches high velocity.

Equations (\ref{cont_simple}) and (\ref{shell_simple}) are not as different
as they first appear.  Both expressions can be written as 
\beq
v_\infty\sim v_{\rm esc}\Gamma^{1/2}
\label{simple}
\eeq
in the limit that $L\gg 4\pi GMc/\kappa$ and $L\gg GMM_{\rm sh}c/R_0^2$, respectively.
But, whereas in the case of a continuous flow the right hand side of equation (\ref{simple})
is evaluated at $R_0$, yielding the result of equation (\ref{cont_simple}),
in the case of a shell the right hand side is evaluated at $R_{\rm UV}$, yielding equation 
(\ref{shell_simple}).  Another way to 
put the difference is that equation (\ref{cont_simple}) implies the gas is accelerated in its
first dynamical time at $r\sim R_0$, whereas equation (\ref{shell_simple}) says that the 
``last" dynamical time at $R_{\rm UV}$ dominates the shell's acceleration and
asymptotic velocity.  

Equation (\ref{shell_simple}) also applies in the case of an optically-thick cloud
that subtends a fraction of the solid angle from the source, with 
the modification that $L/c\rightarrow (L/c)(\pi R_c^2/4\pi r^2)$, where $R_c$ is the cloud radius.
In the special case $R_c\propto r$, which in general does not obtain, the dynamics of shells 
maps trivially to clouds because the cloud solid angle is constant.  
However, even in the simplest case of a cloud expanding into vacuum, a 
new timescale enters the problem, the cloud expansion 
timescale $t_{\rm exp}\sim(d\ln R_c/dt)^{-1} = R_c/c_s$, where $c_s$ is the internal cloud
sound speed and $R_c$ is not a power-law in radius.  
This makes the dynamics of clouds somewhat more complicated than 
shells since the expansion rate of a cloud and
hence the time evolution of its column density is uncoupled from its radial evolution away from the 
source.  The latter is not true for a shell where the radius of the shell is directly coupled to its column density.

In the case of shells, an analogous point, but without the associated dust physics, 
is made in \cite{king03,king05} for the case of shells driven by an AGN wind
and by \cite{dijkstra_loeb08,dijkstra_loeb09} in the case of Ly$\alpha$ scattering.
The case of dusty cloud dynamics in radiation pressure driven galactic winds has been discussed
by \cite{mqt,mmt}.

In this paper, we explore the acceleration of dusty shells and clouds in more detail and apply it to several 
physical systems.  A shell geometry is motivated in some cases by observation of shells in the
massive star and GMC contexts, by detached blue-shifted absorption line profiles 
in the case of some rapidly star-forming galaxies and AGN, and by theoretical
arguments and modeling (e.g., \citealt{yeh_matzner}).
The key point is that dusty shell-like and cloudy outflows can attain significantly
higher velocity then one might guess from an incorrect application of equation (\ref{cont_simple})
as a result of the long phase of acceleration in the single-scattering limit
(eq.~\ref{shell_simple}).  This issue is of particular
importance in the extended gravitational potential wells of galaxies since the 
velocity attained near the source is crucial in determining whether or not it 
will escape to the scale of the virial radius, or, if it falls back, on what timescale.  We
are particularly motivated by the recent discoveries of very fast outflows from post-starburst 
galaxies by \cite{tremonti} and \cite{diamond} (see also \citealt{sell,geach}).

In Section \ref{section:dynamics} we first consider the dynamics of a
shell surrounding a point mass, and then treat extended mass distributions,
as is more appropriate for the dynamics in a galactic gravitational potential.
In Section \ref{section:discussion} we provide a discussion of our results, including
a discussion of fast outflows from galaxies and AGN, the total asymptotic momentum
of radiation pressure accelerated shells and clouds, including
the momentum ratio $\dot{M} v_\infty /(L/c)$ in the shell limit, and we provide 
simple prescriptions for cosmological simulations that
captures the expulsion of gas from rapidly star-forming galaxies.  The extension to 
clouds is treated in Section \ref{section:clouds}.  Our treatment is not as 
extensive as for shells because the dynamics of a given cloud in an outflow depends on 
many parameters including the ensemble of clouds between a given cloud and the radiation
sources, the cloud's internal sound speed and its evolution, the pressure of the background medium, which 
determines a cloud's expansion/thermal history, and destruction processes such as the Kelvin-Helmholz 
instability and evaporation.  Nevertheless, we discuss the dynamics of individual clouds in the context of our
results for the case of shells and show that our primary conclusion --- that fast 
outflows with $v_\infty$ significantly larger than $v_{\rm esc}(R_0)$ are generically obtained --- carries over.

\begin{figure*}
\centerline{\includegraphics[width=8.5cm]{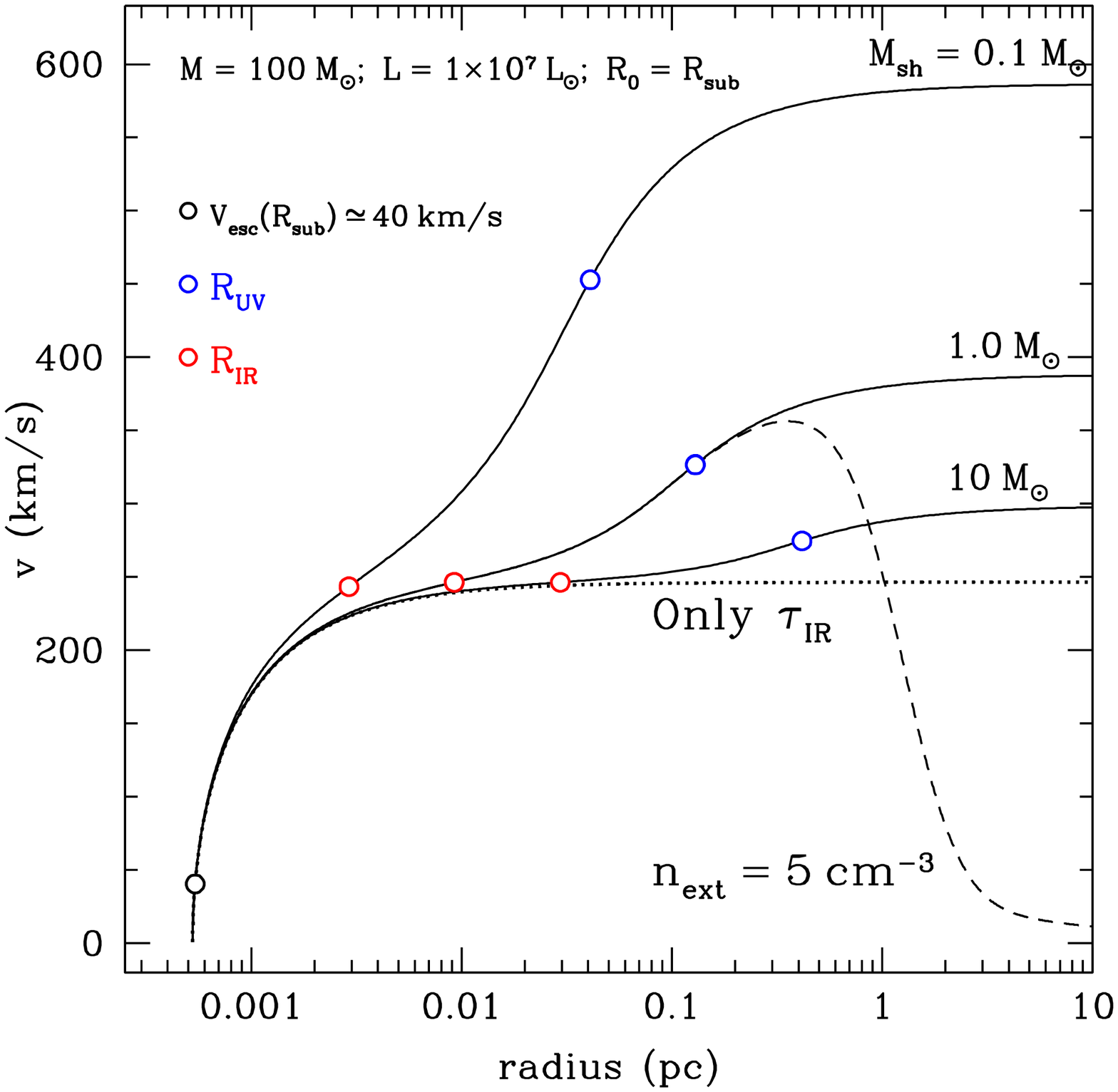}
\includegraphics[width=8.5cm]{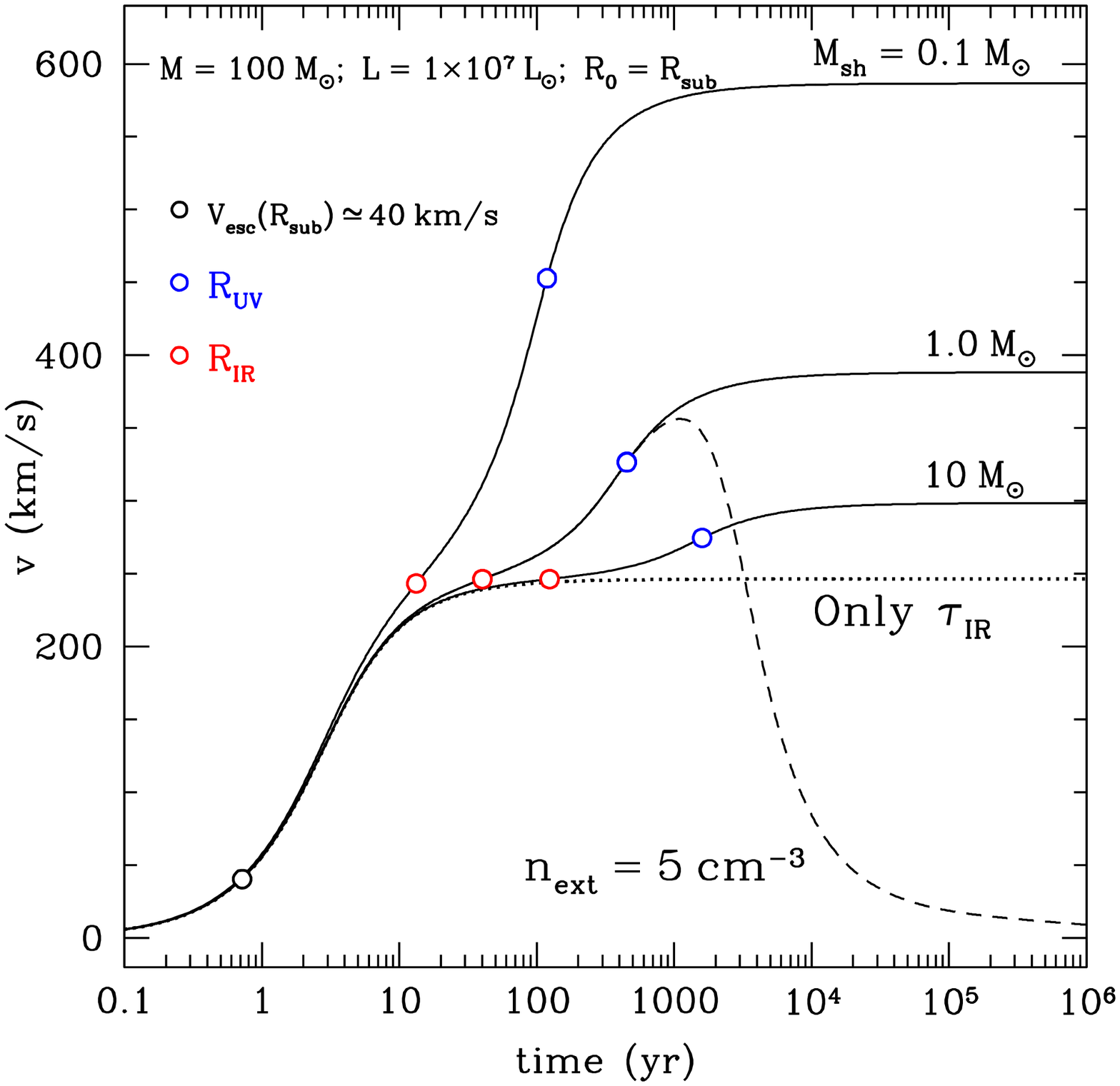}}
\caption{Velocity as a function of radius (left panel) and time (right panel) for dusty shells with mass 10, 1, and 0.1\,M$_\odot$ ejected from a massive star with $M=100$\,M$_\odot$ and $L=10^7$\,L$_\odot$ (solid lines; eq.~\ref{genmom}).  The dotted line shows the evolution if {\it only} the $\tau_{\rm IR}$ term is included in  in equation (\ref{genmom}).  The dashed line shows the dynamics of the 1\,M$_\odot$ shell with a constant density external medium of $n_{\rm ext}=5$\,cm$^{-3}$ (see Section \ref{section:evolve}; eq.~\ref{constantrho}). The black, red, and blue circles denote $v_{\rm esc}$, $R_{\rm IR}$, and $R_{\rm UV}$ (eqs.~\ref{rir}, \ref{ruv}), respectively.  All shells have $R_0=R_{\rm sub}$, the dust sublimation radius, where dust first forms  (see text).}
\label{figure:v}
\end{figure*}

\section{Dynamics of Shells Driven by Radiation Pressure}
\label{section:dynamics}

\subsection{Point Mass}
\label{section:point}

Assume a point source with UV luminosity $L$ and total mass $M$, surrounded by a 
dusty gas shell of mass $M_{\rm sh}$ an initial distance $R_0$ from the central source. 
We define two characteristic radii:
\beqa
R_{\rm IR}&=&\left({\kappa_{\rm IR}\,M_{\rm sh}}/{4\pi}\right)^{1/2}\simeq0.3\,{\rm kpc}\,\kappa_{\rm IR,0.7}^{1/2}M_{g,\,9}^{1/2}\\ \nonumber
R_{\rm IR}&=&\left({\kappa_{\rm IR}\,M_{\rm sh}}/{4\pi}\right)^{1/2}\simeq1.9\times10^3\,{\rm AU}\,\kappa_{\rm IR,0.7}^{1/2}M_{g,\,0}^{1/2}
\label{rir}
\eeqa
is where the shell becomes optically-thin to the re-radiated IR, and 
\beqa
R_{\rm UV}&=&\left({\kappa_{\rm UV}\,M_{\rm sh}}/{4\pi}\right)^{1/2}\simeq4\,{\rm kpc}\,\kappa_{\rm UV,3}^{1/2}M_{g,\,9}^{1/2}\\ \nonumber
R_{\rm UV}&=&\left({\kappa_{\rm UV}\,M_{\rm sh}}/{4\pi}\right)^{1/2}\simeq4.6\times10^4\,{\rm AU}\,\kappa_{\rm UV,3}^{1/2}M_{g,\,0}^{1/2}
\label{ruv}
\eeqa
is where the shell becomes optically-thin to the UV radiation from the source.  In the above,
$M_{{\rm sh},\,x}=M_{\rm sh}/10^x$\,M$_\odot$, we have scaled for both the galaxy and
stellar contexts,
$\kappa_{\rm UV,3}=\kappa_{\rm UV}/10^3\,f_{\rm dg,\,MW}$\,cm$^2$ g$^{-1}$ of gas, where
$f_{\rm dg,\,MW}$ is the dust-to-gas ratio scaled to the Milky Way value.
and $\kappa_{\rm IR,0.7}=\kappa_{\rm IR}/10^{0.7}\,f_{\rm dg,\,MW}$\,cm$^2$ g$^{-1}$
roughly approximates the Rosseland-mean dust opacity over a range of temperatures
from $\sim100-1000$\,K.

The general expression for momentum conservation for a thin shell of mass $M_{\rm sh}$, 
approximately valid in the limits of both small and large UV and IR optical depth is 
\beq
\frac{d}{dt}\left(M_{\rm sh} v\right)=-\frac{GMM_{\rm sh}}{r^2}+\left(1+\tau_{\rm IR}-e^{-\tau_{\rm UV}}\right)\frac{L}{c},
\label{genmom}
\eeq
where 
\beq
\tau_{\rm IR,\,UV}=\kappa_{\rm IR,\,UV}M_{\rm sh}/(4\pi r^2)
\eeq
are the IR and UV optical depths of the shell and where we have assumed that the dust and gas are dynamically
coupled.  Note the three terms multiplying $L/c$ in equation 
(\ref{genmom}).  The first term (``1") is due to the direct radiation field.  
It represents the radiation pressure force exerted if each photon interacts just once with the dusty medium,
is converted into an IR photon, and then escapes the system.
The second term 
accounts for reprocessed radiation if the shell is optically-thick to the re-radiated IR emission.
The third term goes to the appropriate limit when $\tau_{\rm UV}\ll1$ ($\tau_{\rm IR}\ll1$ also), 
canceling the ``1" and yielding the familiar optically-thin radiation pressure force for a source
dominated by UV emission ($\kappa_{\rm UV}L/4\pi r^2 c$).  

Setting the acceleration in equation (\ref{genmom}) equal to zero, we obtain the 
generalized Eddington limit for a shell starting at $R_0$:
\beq
L_{\rm Edd}=\frac{GMM_{\rm sh} c}{R_0^2}\left(1+\tau_{\rm IR}-e^{-\tau_{\rm UV}}\right)^{-1},
\label{ledd}
\eeq
where $M$ and $M_{\rm sh}$ are understood to be the total mass interior to $R_0$ and  
the shell mass at $R_0$, respectively.  The Eddington ratio is then
\beq
\Gamma_{\rm tot}=L/L_{\rm Edd}.
\label{genedd}
\eeq
There are three characteristic Eddington ratios, depending on 
the optical depth of the shell at $R_0$. If the shell is optically-thick
to the IR ($\tau_{\rm IR}(R_0)>1$), then the relevant Eddington ratio is 
\beq
\Gamma_{\rm IR}=L/(4\pi G M c/\kappa_{\rm IR}).
\label{gammair}
\eeq
If the shell is optically-thin to the IR, but optically-thick to the UV ($\tau_{\rm IR}<1$, $\tau_{\rm UV}>1$) at $R_0$, then the single-scattering Eddington ratio (the ``1" in eqs.~\ref{genmom} and \ref{genedd}) 
\beq
\Gamma_{\rm SS}=L/(GMM_{\rm sh} c/R_0^2),
\label{gammass}
\eeq
determines the dynamics. Finally, if the shell is optically-thin to the incident UV
radiation ($\tau_{\rm UV}<1$ at $R_0$), 
then the relevant Eddington ratio is 
\beq
\Gamma_{\rm UV}=L/(4\pi G M c/\kappa_{\rm UV}).
\label{gammauv}
\eeq
Before solving equation (\ref{genmom}) for specific example systems, 
we consider several simple analytic limits.

We first assume that the mass of the shell is constant as a function of radius
(i.e., the shell expands into vacuum).
Then, dropping the $e^{-\tau_{\rm UV}}$ term in equation (\ref{genmom})
 in the regime $R_0\leq r\leq R_{\rm UV}$ ($\tau_{\rm UV}>1$) and solving for 
 the velocity at $v_{\rm UV}=v(R_{\rm UV})$, one finds that
\beq
v_{\rm UV}^2=v_0^2+\frac{2GM}{R_0}\left(\Gamma_{\rm SS}\frac{R_{\rm UV}}{R_0}+\Gamma_{\rm IR}-1\right)\left(1-\frac{R_0}{R_{\rm UV}}\right),
\label{vuv}
\eeq
where $v_0=v(R_0)$.
Once the shell reaches $R_{\rm UV}$ it becomes optically-thin to the incident UV photons, and 
the momentum equation for the shell in the regime $R_{\rm UV}\leq r\leq \infty$ is just
\beq
v\,\frac{dv}{dr}=-\frac{GM}{r^2}+\frac{\kappa_{\rm UV} L}{4\pi r^2 c}.
\eeq
Solving, and substituting from equation (\ref{vuv}) one finds that 
\beqa
v_\infty^2=v_0^2&+&\frac{2GM}{R_0}
\left[ \Gamma_{\rm SS}\frac{R_{\rm UV}}{R_0}+\Gamma_{\rm IR}-1\right]
\left(1-\frac{R_0}{R_{\rm UV}}\right) \nonumber \\
&+&\frac{GM}{R_{\rm UV}}\left(\Gamma_{\rm UV}-1\right)
\label{vinf0}
\eeqa
which can be rewritten as
\beqa
v_\infty^2=v_0^2&+&\frac{2GM}{R_0}
\left[ 2\Gamma_{\rm SS}\frac{R_{\rm UV}}{R_0}\left(1-\frac{R_0}{2R_{\rm UV}}\right) \right. \nonumber \\
&\,&\hspace*{1.5cm}\left.+\,\Gamma_{\rm IR}\left(1-\frac{R_0}{R_{\rm UV}}\right)-1\right]
\label{vinf}
\eeqa
These expressions are important.
The ratio  $R_{\rm UV}/R_0$ can be 
much larger than unity and thus, even if the initial Eddington ratio of the 
flow is equal to unity $\Gamma_{\rm tot}\simeq1$ at $R_0$, 
the shell can still be accelerated to $v_\infty\gg v_{\rm esc}$ if it starts with $\tau_{\rm UV}>1$.
Taking $R_0\ll R_{\rm UV}$, note that the ratio of the first two terms in square brackets in equation (\ref{vinf}) is 
\beq
\left(2\Gamma_{\rm SS}\frac{R_{\rm UV}}{R_0}\right)\,\frac{1}{\Gamma_{\rm IR}}=
\frac{2}{\tau_{\rm IR}(R_0)^{1/2}}\left(\frac{\kappa_{\rm UV}}{\kappa_{\rm IR}}\right)^{1/2},
\eeq
and thus the $\Gamma_{\rm SS}$ 
term dominates unless $\tau_{\rm IR}(R_0)>4\kappa_{\rm UV}/\kappa_{\rm IR}\sim10^3$.
Thus, assuming $\Gamma_{\rm IR}\ll 2\Gamma_{\rm SS}R_{\rm UV}/R_0$ and then 
taking $v_0\sim0$, $R_0\ll R_{\rm UV}$ in equation (\ref{vinf}), one finds that
\beq
v_\infty\simeq v_{\rm esc} \left(2\Gamma_{\rm SS}\frac{R_{\rm UV}}{R_0}\right)^{1/2} 
=\,v_{\rm esc} \Gamma_{\rm SS}^{1/2}\left[4\,\tau_{\rm UV_0}\right]^{1/4},
\label{vtau}
\eeq
where $v_{\rm esc}$ is understood to be the escape velocity from the launch radius $R_0$
in a point-mass potential, and $\tau_{\rm UV_0}=\tau_{\rm UV}(R_0)$ is the UV optical 
depth at $R_0$.  This expression can be rewritten in terms of the luminosity  ($L_x=L/10^x$\,L$_\odot$) as\footnote{Note
that these expressions are only applicable when $R_0\ll R_{\rm UV}$.  As the mass of the shell 
becomes small and $\tau_{\rm UV}(R_0)$ becomes less than unity, $v_\infty=v_{\rm esc}(R_0)(\Gamma_{\rm UV}(R_0)-1)^{1/2}$
(see eq.~\ref{vmax_thin}).}
\beqa
v_\infty&\simeq&\left(\frac{4 R_{\rm UV}L}{M_{\rm sh} c}\right)^{1/2}  
\simeq\left(\frac{2L}{c}\right)^{1/2}\left(\frac{\kappa_{\rm UV}}{\pi M_{\rm sh}}\right)^{1/4} \nonumber \\
&\simeq&320\,{\rm km\,\,s^{-1}}\,L_{7}^{1/2}\kappa_{\rm UV,\,3}^{1/4}\,M_{\rm sh,\,0}^{-1/4},\nonumber \\
&\simeq&1800\,{\rm km\,\,s^{-1}}\,L_{13}^{1/2}\kappa_{\rm UV,\,3}^{1/4}\,M_{\rm sh,\,9}^{-1/4}.
\label{vx}
\eeqa

As discussed in Section \ref{section:introduction},
equation (\ref{vinf}), and the simple results given in equations (\ref{vtau}) and (\ref{vx}) 
for the asymptotic velocity of a radiation pressure driven shell are 
qualitatively and quantitatively different from the expectation that 
$v_\infty \simeq v_{\rm esc}(R_0) (\Gamma_{\rm tot}-1)^{1/2}$ (eq.~\ref{cont_simple}). 
In particular, from equation (\ref{vtau}) one sees that $v_\infty$ can significantly exceed
$v_{\rm esc}(R_0)$, even for an Eddington ratio near unity at $R_0$: high Eddington ratios are not
required for high velocities with respect to the escape velocity.  Instead, the initial 
value of the UV optical depth through the shell determines its dynamical evolution.
The typical ``boost'' in the asymptotic velocity compared to $v_{\rm esc}(R_0)$ is
\beqa
\Gamma_{\rm SS}^{1/2}\left[4\,\tau_{\rm UV_0}\right]^{1/4}
&\simeq&9\,\,\Gamma_{\rm SS}^{1/2}R_{0,\,\rm 0.1\,kpc}^{-1/2}\,\kappa_3^{1/4}\,M_{g,\,9}^{1/4} \nonumber \\ 
&\simeq&73\,\,\Gamma_{\rm SS}^{1/2}R_{0,\,\rm 10AU}^{-1/2}\kappa_3^{1/4}\,M_{g,\,0}^{1/4}
\label{boost}
\eeqa
Thus, in the context of galactic winds, even if $\Gamma_{\rm SS}\sim1$, one expects the 
asymptotic velocity of the shell to exceed the escape velocity from the launch region 
by nearly an order of magnitude (see Sections \ref{section:extended} and \ref{section:evolve} for a discussion of 
extended galactic potentials).  In the stellar context, the boost may be larger.
The physics of this enhancement in the asymptotic velocity comes simply from the 
radial dependence of the single-scattering radiation pressure term; in particular, aside
from acting only until $r\simeq R_{\rm UV}$ it has no radial fall-off, whereas both the gravitational
acceleration and the flux drop with radius as $r^{-2}$.  The result that  $v_\infty^2 /R_{\rm UV} \sim L/M_{\rm sh}c$
is precisely what one would then get from dimensional analysis of equation (\ref{genmom}), which 
is equivalent to the result $v_\infty \sim v_{\rm esc}\Gamma^{1/2}$, {\it but evaluated at $R_{\rm UV}$} (see 
discussion after eq.~\ref{shell_simple}).

As an aside, note that the dynamical stability of a dusty shell 
to radial perturbations is different depending 
on whether or not the radiation pressure force is 
$f_{\rm rad}=\kappa L/4\pi r^2 c$ or $L/M_{\rm sh} c$.  In
the latter, single-scattering limit, shells are unstable to radial perturbations since $f_{\rm rad}\propto r^0$,
whereas the gravitational force is $f_{\rm grav}\propto r^{-2}$.  Thus, if an equilibrium is established with 
$\Gamma_{\rm SS}=1$ small
perturbations would drive the shell to smaller $r$ causing collapse, or larger $r$, causing dynamical
escape with $v_{\infty}$ given by equation (\ref{vx}), different from the behavior if
 $f_{\rm rad}\propto r^{-2}$.

Figure \ref{figure:v} shows integrations of equation (\ref{genmom}) for a massive star outburst
(left panel).  Shells of 0.1, 1, and 10\,M$_\odot$
are accelerated from the dust sublimation radius $R_{\rm sub}=(L/16\pi \sigma_{\rm sb}T_{\rm sub}^4)^{1/2}$,
where $T_{\rm sub}\simeq1500$\,K to pc scales for an outburst from a massive star with $M=100$\,M$_\odot$
and luminosity $10^7$\,L$_\odot$.  The solid lines show the full solution to equation (\ref{genmom}),
whereas the dotted line shows the solution with {\it only} the $\tau_{\rm IR}$ term.  The dashed line, which shows the change in dynamics when the 1\,M$_\odot$ shell interacts with a constant density medium, is discussed in Section \ref{section:evolve}. The black, red, and blue dots mark the radial location of $v_{\rm esc}(R_{\rm sub})$,  $R_{\rm IR}$ and $R_{\rm UV}$ respectively.  The initial IR optical depths of the shells are $\simeq30$, 300, and 3000,  and $\Gamma_{\rm SS}\simeq1$, 0.1, and 0.01, respectively (see \ref{section:uncertain}). For all shells, $\Gamma_{\rm IR}\simeq40$.  

Based on the fact that the initial IR optical depth is larger than unity at $R_0=R_{\rm sub}(\simeq100$\,AU) and that $\Gamma_{\rm IR}\gg\Gamma_{\rm SS}$, one might have expected that   $v_\infty\simeq v_{\rm esc}(R_{\rm sub})\left(\Gamma_{\rm IR}-1\right)^{1/2}\simeq250$\,km s$^{-1}$, as shown by the dotted line, which only includes the $\tau_{\rm IR}$ term in equation (\ref{genmom}).  Instead, the single-scattering term in equation (\ref{genmom}) dominates the dynamics on large scales, accelerating the shells to velocities much larger than $v_{\rm esc}(R_{\rm sub})\simeq40$\,km s$^{-1}$ since $R_{\rm UV}/R_{\rm 0}\gg1$.  

If we were to instead drop the $\tau_{\rm IR}$ term from our solution of equation (\ref{genmom}) for the shells in Figure \ref{figure:v}, the initial rapid rise in velocity to $\simeq250$\,km/s is missing for the 0.1\,M$_\odot$ shell, but its subsequent dynamics is similar to that shown because $\Gamma_{\rm SS}$ is just larger than unity at $R_0$.  For the other more massive shells, $\Gamma_{\rm SS}$ is less than unity at $R_0$ and they fall back toward the central star, even though on larger scales in the full solution shown in Figure \ref{figure:v} this term is responsible for substantial acceleration.  Finally, if we were to assume incorrectly that the shells were optically-thin to the UV at $R_0$, so that the radiation pressure term on the right hand side of equation (\ref{genmom}) was just $\tau_{\rm UV}L/c$, the shells would be accelerated rapidly to $v_{\rm esc}(R_0)\Gamma_{\rm UV}^{1/2}\sim3600$\,km/s.  These various limits of the momentum equation are useful in considering the dynamics of rapidly expanding clouds discussed in Section \ref{section:clouds}.

Turning back to the solid lines in Figure \ref{figure:v}, one might have also expected that the shell with the highest mass and highest initial optical depth to  have the highest asymptotic velocity, but because these results are shown for fixed luminosity,  and because $\Gamma_{\rm SS}\propto M_{\rm sh}^{-1}$, one has that $v_\infty\propto M_{\rm sh}^{-1/4}$ (eq.\ \ref{vx}). Thus, low mass shells are driven to higher asymptotic velocity than higher mass shells at fixed $L$; for $M_{\rm sh}=0.01$\,M$_\odot$ (not shown), $v_\infty\simeq970$\,km s$^{-1}$.  More discussion of these types of eruptions are included  in Section \ref{section:massive}.  Since such outbursts have high initial IR optical depths, their dynamics may be important in assessing multi-dimensional instabilities that  may limit the momentum coupling between the radiation field and the shell as it is accelerated (see \ref{section:uncertain}).\footnote{Note that throughout this work we assume that $Lt/(M_{\rm sh} v^2/2)\gg1$.   For high enough IR optical depths it is possible for the radiation field to do sufficient work on the matter to  enter the ``photon tiring'' regime discussed by \cite{owocki_gayley} and \cite{owocki} in the context of line-driven winds. Examining $Lt/(M_{\rm sh} v^2/2)$ as a function of radius for the models in Figure \ref{figure:v}, we find that  it is always larger than unity, although it becomes as low as $\sim2$ on scales smaller than the $\tau_{\rm IR}=1$ point (red circle) for the model with $M_{\rm sh}=10$\,M$_\odot$. An estimate of the critical shell mass such that  $Lt/(M_{\rm sh} v^2/2)=1$ and photon tiring becomes important  is $M_{\rm sh,\,tiring}=L^{3/4}c^{3/2}\pi^{1/4}/(2^{3/2}\sigma_{\rm SB}^{5/4}T_{\rm sub}^5\kappa_{\rm IR}^{3/2})$, obtained by taking $\tau_{\rm IR}(R_{\rm sub})=2c/v_\infty$ in the limit that $\Gamma_{\rm IR}\gg1$ at $R_{\rm sub}$.   For the parameters of Figure \ref{figure:v} this is $M_{\rm sh,\,tiring}\simeq8\,{\rm M_\odot}\,L_7^{3/4}T_{1500}^{-5}\kappa_{\rm IR,\,0.7}^{-3/2}$. Note the  strong dependencies. \label{tiring}}

Finally, note that the characteristic acceleration timescale $t_{\rm acc}\sim v_\infty/(dv/dt)$ 
is long on the scale of observations of 
a single massive star outburst:
\beqa
t_{\rm acc}&\sim&\left(\frac{2c}{L}\right)^{1/2}\left(\frac{\kappa_{\rm UV}}{4\pi}\right)^{1/4}M_{\rm sh}^{3/4}  \nonumber \\
&\sim& 1100\,\,{\rm yr} \,\,\,L_7^{-1/2} M_{\rm sh,\,0}^{3/4} \kappa_{\rm UV,\,3}^{1/4} \nonumber \\
&\sim& 6.4\times10^6\,\,{\rm yr} \,\,\,L_{13}^{-1/2} M_{\rm sh,\,9}^{3/4} \kappa_{\rm UV,\,3}^{1/4}.
\label{tacc}
\eeqa
The scaling for the massive star outburst agrees with the calculations shown in 
Figure \ref{figure:v} and implies that high-velocity shells driven by this physics will be associated
with many old outbursts.

\begin{figure*}
\centerline{\includegraphics[width=8.5cm]{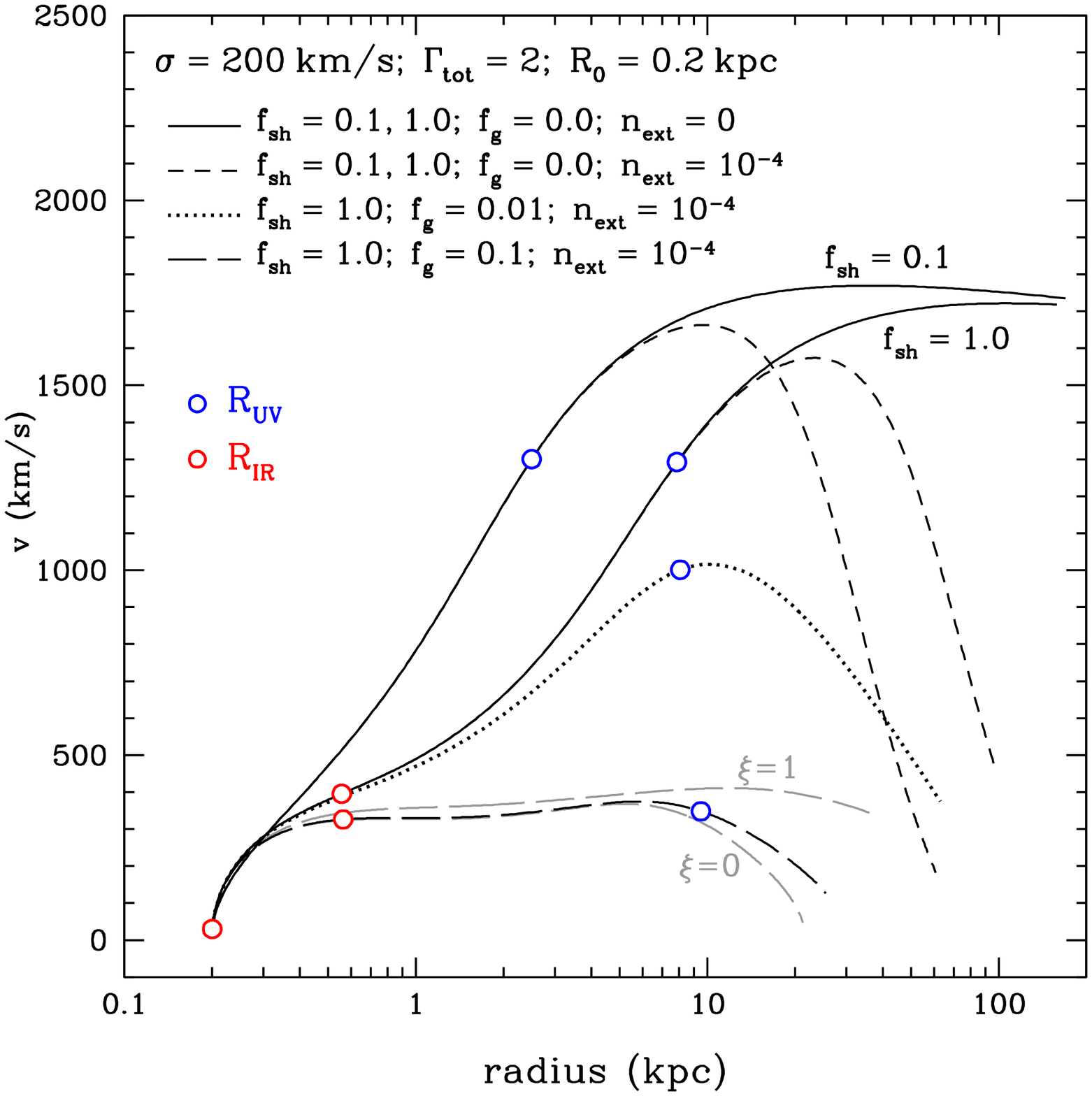}
\includegraphics[width=8.5cm]{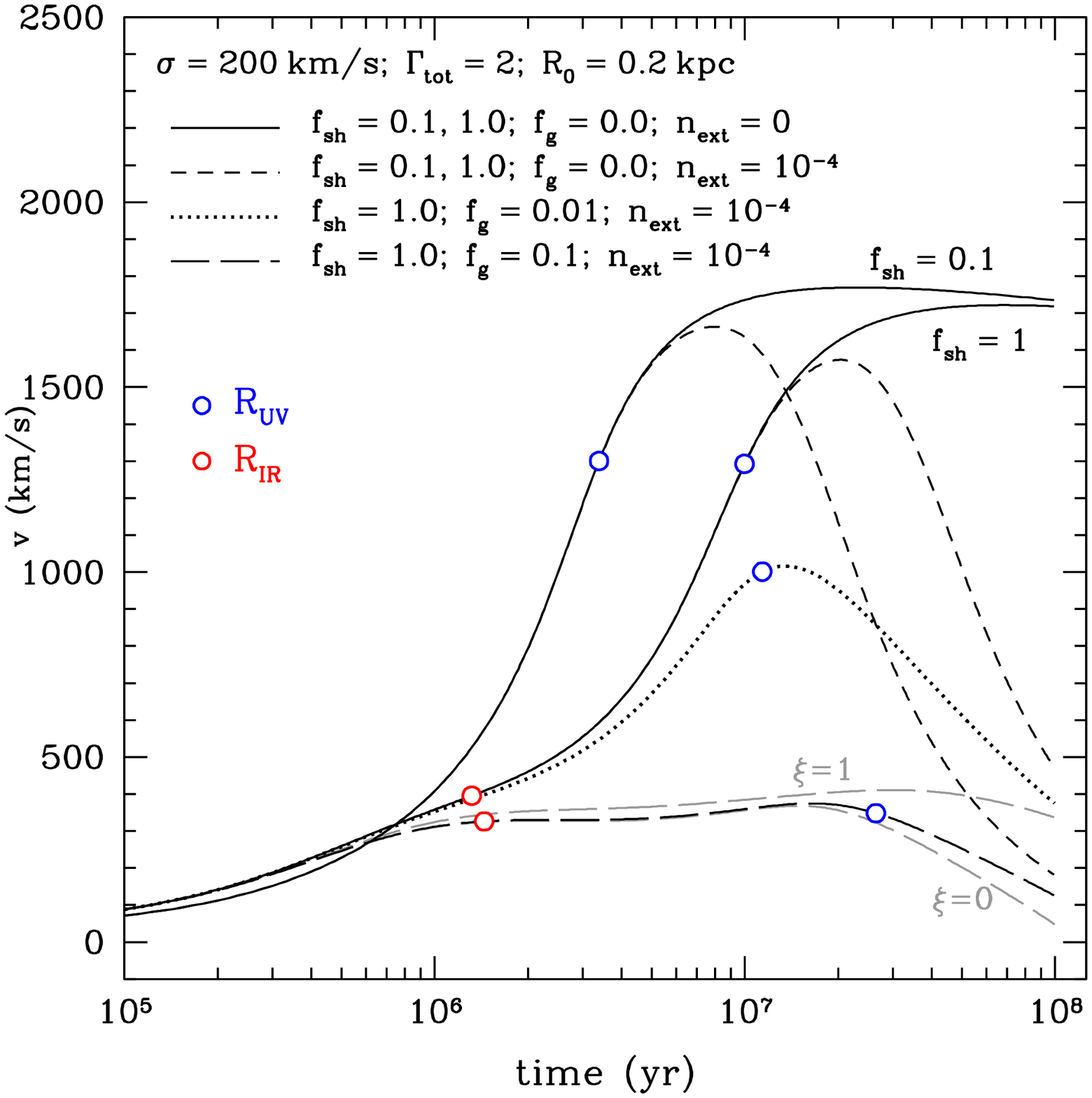}}
\caption{Velocity as a function of radius (left panel) and time (right panel) 
for dusty shells in an isothermal potential with $\sigma=200$\,km s$^{-1}$, starting from 
a launch radius of $R_0=0.2$\,kpc and $\Gamma_{\rm tot}=2$ (eqs.~\ref{ledd} \& \ref{genedd}).
The two solid lines show the evolution for a freely expanding shell with no external medium ($f_g=0$, $n_{\rm ext}=0$\,cm$^{-3}$) and with $f_{\rm sh}=0.1$, and 1, where $f_{\rm sh}$ is defined by equation (\ref{fsh}),
so that $M_{\rm sh}\simeq3.7\times10^8$\,M$_\odot$ and $\simeq3.7\times10^9$\,M$_\odot$, respectively.
Given $\Gamma_{\rm tot}$, the total luminosities are $\simeq8.9\times10^{12}$\,L$_\odot$ and $\simeq2.6\times10^{13}$\,L$_\odot$ for $f_{\rm sh}=0.1$, and 1, respectively. The short dashed lines show the evolution of both shells if they interact with a constant density halo that has $n_{\rm ext}=10^{-4}$\,cm$^{-3}$ (see Section \ref{section:evolve}; eq.~\ref{constantrho}).  The dotted and long dashed lines show the evolution of the $f_{\rm sh}=1$ shell if we add an external gas density distribution that follows equation (\ref{isodensity}) with $f_g=0.01$ and 0.1, respectively (see eq.~\ref{fgcrit}).  The dust content of the swept up gas in all cases is assumed equivalent to a fraction $\xi=0.1$ (see eq.~\ref{xi}) times the Milky Way value, except for the two gray long-dashed curves, which show the change in the dynamics for the $f_{\rm sh}=1.0$, $f_g=0.1$, and $n_{\rm ext}=10^{-4}$\,cm$^{-3}$ model if the external medium has zero dust ($\xi=0$) or a Milky Way value ($\xi=1$).   The red, and blue circles denote  $R_{\rm IR}$ and  $R_{\rm UV}$ (eqs.~\ref{rir}, \ref{ruv}), respectively. See Fig.~\ref{figure:ratio} for the momentum ratio $\zeta$ (Section \ref{section:momentum}) and column density evolution of these models.}
\label{figure:viso}
\end{figure*}

\subsection{Extended Mass Distributions with Fixed Shell Mass}
\label{section:extended}

In the context of outflows driven from galaxies, it is important to consider the
extended stellar and dark matter potentials.
For illustration we assume that the total mass distribution is an isothermal sphere:
$M(r)=2 \sigma^2 r/G$, where $\sigma$ is the velocity dispersion.  Assuming momentarily 
that the  mass of the shell is constant, we then have that 
\beq
M_{\rm sh}\,v\,\frac{dv}{dr}=-\frac{2\sigma^2}{r}+\left(1+\tau_{\rm IR}-e^{-\tau_{\rm UV}}\right)\frac{L}{c}.
\label{genmom2}
\eeq
and
\beq
L_{\rm Edd}=\frac{2\sigma^2M_{\rm sh} c}{R_0}\left(1+\tau_{\rm IR}-e^{-\tau_{\rm UV}}\right)^{-1}.
\label{leddiso}
\eeq
We again define 
\beq
\Gamma_{\rm tot}=L/L_{\rm Edd},
\label{gammatotiso}
\eeq
and the three characteristic Eddington ratios
\beq
\Gamma_{\rm IR, \, UV}= L/(8\pi \sigma^2 R_0 c/\kappa_{\rm IR,\,UV})
\eeq
and
\beq
\Gamma_{\rm SS}=L/(2\sigma^2M_{\rm sh} c/R_0).
\eeq
Again taking $e^{-\tau_{\rm UV}}\sim0$ in the regime $R_0\leq r\leq R_{\rm UV}$, we find that 
\beqa
v_{\rm UV}^2&=&v_0^2+4\sigma^2\left[\left(\Gamma_{\rm SS}\frac{R_{\rm UV}}{R_0}+
\Gamma_{\rm IR}\right)\left(1-\frac{R_0}{R_{\rm UV}}\right)\right. \\ \nonumber
&\,&\hspace*{2.8cm}-\left.\ln\left(\frac{R_{\rm UV}}{R_0}\right)\right],
\label{vuve}
\eeqa
and then integrating to an outer radius $R_{\rm out}$ where the approximation of 
an isothermal potential breaks down,
\beqa
v_{\rm out}^2&=&v_0^2+4\sigma^2
\left[2\Gamma_{\rm SS}\frac{R_{\rm UV}}{R_0}\left(1-\frac{R_0}{R_{\rm UV}}-\frac{R_{\rm UV}}{2 R_{\rm out}}\right)\right. \nonumber \\
&\,&\hspace*{1.3cm}\left.+\Gamma_{\rm IR}\left(1-\frac{R_0}{R_{\rm UV}}\right)
-\ln\left(\frac{R_{\rm out}}{R_0}\right)\right].
\label{voutiso}
\eeqa
The same factor that appears in the point-mass limit ---$(2\Gamma_{\rm SS}R_{\rm UV}/R_0)^{1/2}$ --- 
which can boost the asymptotic velocity far above $v_{\rm esc}(R_0)$ (eq.~\ref{vx}) also appears in the 
limit of an extended mass distribution.  Assuming that $\Gamma_{\rm tot}\ge1$, $R_0\ll R_{\rm UV}$,
$R_{\rm UV}\ll R_{\rm out}$, and that $\Gamma_{\rm IR}\ll 2\Gamma_{\rm SS} R_{\rm UV}/R_0$,
one finds that 
\beqa
v_{\rm out}&\simeq& 2\sigma\Gamma_{\rm SS}^{1/2}\left[4\tau_{\rm UV_0}\right]^{1/4} \nonumber \\
&\simeq&18\sigma\,\,\Gamma_{\rm SS}^{1/2}R_{0,\,\rm 0.1\,kpc}^{-1/2}\,\kappa_{\rm UV,3}^{1/4}\,M_{g,\,9}^{1/4}
\label{viso}
\eeqa
fully analogous with equation (\ref{vx}) in the point-mass limit. If we define $f_{\rm sh}$ as the fraction
of mass within the launch radius $R_0$ that goes into the shell,
\beq
M_{\rm sh}=f_{\rm sh}\,\frac{2\sigma^2 R_0}{G},
\label{fsh}
\eeq
we find that 
\beqa
v_{\rm out}&\simeq&2\Gamma_{\rm SS}^{1/2}\left(\frac{2\,\kappa_{\rm UV}\,\sigma^6\,f_{\rm sh}}{\pi R_0 G}\right)^{1/4} \nonumber \\
&\simeq&4\times10^3\,{\rm km\,s^{-1}}\,\Gamma_{\rm SS}^{1/2}\kappa_{\rm UV,\,3}^{1/4}f_{\rm sh}^{1/4}R_{0,\,\rm0.1\,kpc}^{-1/4}.
\eeqa
Note that these analytic estimates overestimate $v_{\rm out}$ because they ignore the logarithmic factor in equation (\ref{voutiso}).

Figure \ref{figure:viso} shows the velocity evolution of massive dusty shells launched in an
isothermal potential.  We assume this potential extends to 100\,kpc for simplicity, even though  this approximation breaks down for real galaxies on the scale of $\sim10$s of kpc.  For massive shells (e.g., $f_{\rm sh}=1$), we include the self-gravity of the shell itself in the total mass $M$ in our solution to the momentum equation (\ref{genmom}) using $M=M(<r)+M_{\rm sh}/2$. The two solids lines show velocity as a function of radius (left panel) and time (right panel) for $f_{\rm sh}=0.1$ and $f_{\rm sh}=1$ (see eq.~\ref{fsh}), corresponding to $M_{\rm sh}=3.7\times10^8$ and $3.7\times10^9$\,M$_\odot$, respectively, launched from a galaxy with $\sigma=200$\,km s$^{-1}$, and from a radius $R_0=0.2$\,kpc.  For each shell, we  assume $\Gamma_{\rm tot}=2$, so that $L\simeq8.9\times10^{12}$, $2.6\times10^{13}$\,L$_\odot$ as might be provided by a  central starburst and/or active galactic nucleus.  These two models (solid lines) assume that the shell expands into vacuum.  In Section \ref{section:evolve} we calculate the dynamics of shells expanding into a constant density medium and a medium with an isothermal sphere density profile (dashed and dotted lines), and with a varying gas-to-dust ratio for the swept up medium (gray dashed lines).  See Section \ref{section:evolve}.
   
For the parameters of the shells shown by the solid lines in Figure \ref{figure:viso},  the naive expectation in many models of galactic winds would have been that $v_\infty\simeq 2\sigma(\Gamma_{\rm tot}-1)^{1/2}\simeq400$\,km s$^{-1}$.  However,  because of the factor $(2\Gamma_{\rm SS}R_{\rm UV}/R_0)^{1/2}$ in equation (\ref{voutiso}) caused by long-term acceleration of the shell in the single-scattering limit, the actual velocities are $\sim4$ times this value, reaching $\sim1600$\,km s$^{-1}$ on $\sim1-10$\,kpc scales.  These high velocities have important implications for observations (see Section \ref{section:discussion}).

\subsection{Evolving Shells}
\label{section:evolve}

In the general case of a massive shell driven into the circumgalactic medium of 
highly star-forming galaxies, we expect the shell to sweep up mass and the assumptions
of the previous section break down.  In the limiting case that the shell sweeps up less 
than its initial mass by the time it reaches $R_{\rm UV}$, we expect the dynamics to 
be qualitatively similar.  However, if the mass of the swept-up material approaches 
the initial mass of the shell on the scale of $R_{\rm UV}$,
we expect the shell dynamics to be altered.  If we assume that the circumgalactic 
gas takes the form of a static isothermal sphere with gas density 
\beq
\rho=\frac{f_g\,\sigma^2}{2\pi G r^2},
\label{isodensity}
\eeq
the critical value for $f_g$ such that the swept up gas mass $M_{\rm sh}(R_{\rm UV})$ is equal to the 
initial shell mass $M_{\rm sh}(R_0)$ is 
\beq
f_{g, \, \rm crit}=\frac{G}{\sigma^2}\left(\frac{\pi M_{\rm sh}(R_0)}{\kappa_{\rm UV}}\right)^{1/2}
\simeq 0.01\sigma_{200}^{-2}\,M_{\rm sh,\,9}^{1/2}\,\kappa_{\rm UV,\,3}^{-1/2}.
\label{fgcrit}
\eeq
Thus, for $f_g\gtrsim f_{g, \, \rm crit}$, we expect the shell dynamics to 
be different from the solid lines shown in Figure \ref{figure:viso}.  In particular, 
we expect the shell to decelerate, in accord with momentum conservation.  

An analogous estimate can be made in the stellar case, where the shell from the 
eruption sweeps up the matter in a steady preceding stellar wind of mass loss rate $\dot{M}$.  
In this case, the density profile is an isothermal sphere with $\rho=\dot{M}/(4\pi r^2 v_w)$, where 
$v_w$ is the wind velocity.  Setting the total swept up mass $\dot{M} (r-R_0)/v_w$ equal to the 
initial shell mass, one derives a critical mass loss rate such that the shell sweeps
up its own mass on a scale $R_{\rm UV}$:
\beqa
\dot{M}_{\rm crit}&=&\left(\frac{4\pi M_{\rm sh}v_w^2}{\kappa_{\rm UV}}\right)^{1/2} \nonumber \\
&\simeq&8\times10^{-4}\,{\rm M_\odot\,\,yr^{-1}}M_{\rm sh,\,0}^{1/2}\,\kappa_{\rm UV,\,3}^{-1/2}\,v_{w,\,100},
\label{windrho}
\eeqa
where $v_{w,\,100}=v_w/100$\,km s$^{-1}$.\footnote{For 
the purposes of this estimate we include $v_w$ in the wind mass profile, 
but neglect the bulk flow of the wind matter in the momentum equation 
in calculating $\dot{M}_{\rm crit}$.
}

If we take the surrounding 
external medium to have constant density, we can estimate the critical density 
such that the swept up mass is equal to the initial shell mass at $R_{\rm UV}$.  This
limit is applicable to massive stars in constant density circumstellar envelopes,
or shells interacting with the surrounding ISM, and to shells driven from galaxies
that sweep up matter from the hot near-constant density halo (e.g., \citealt{maller_bullock}).
The critical external density required to slow the shell on a scale $R_{\rm UV}$ is
\beqa
n_{\rm ext, \,crit}&=&4.5\times10^{3}\,\,{\rm cm^{-3}}\,\kappa_{\rm UV,3}^{-3/2}\,M_{\rm sh,0}^{1/2} \nonumber \\
&=&0.1\,\,{\rm cm^{-3}}\,\kappa_{\rm UV,3}^{-3/2}\,M_{\rm sh,9}^{1/2}.
\label{constantrho}
\eeqa
The normalization for the massive star case in the first line is very large compared to what would be expected in the 
average region of a star-forming galaxy like the Milky Way (e.g., $\sim1$\,cm$^{-3}$), and even larger than an average volume within most starburst galaxies (e.g., \citealt{krumholz_thompson07}).  The normalization for the 
galaxy case in the second line is also large compared to the very large-scale approximately constant density hot component of the circumgalactic medium, which may have $n\sim10^{-3}-10^{-4}$\,cm$^{-3}$ (e.g., \citealt{maller_bullock}). One might then expect that in both cases the constant density medium will not 
have an important effect on the shell dynamics.  
However, because the swept up mass scales with $r^3$ in the constant density case we 
expect the shell to slow significantly on scales larger than $R_{\rm UV}$ even for 
$n_{\rm ext}\ll n_{\rm ext,\,crit}$, but that the maximum velocity of the shell at $R_{\rm UV}$ will
not be much smaller than the estimates above as long as
$n_{\rm ext}$ is not greater than $n_{\rm ext,\,crit}$.  

As an example, the dashed line in Figure \ref{figure:v} shows the evolution of 
a 1\,M$_\odot$ shell driven into a constant density medium with $n_{\rm ext}=5$\,cm$^{-3}$.
Of course, the circumstellar medium around a massive star in outburst is likely to 
have a complex density structure with a wind-blown bubble, but equations (\ref{windrho}) and 
(\ref{constantrho}) show that the density must be very high to slow the shell on
scales much smaller than $R_{\rm UV}$.

Similar examples of a shell interacting with a medium, but in the galactic case,
are shown by the dashed and dotted lines in Figure \ref{figure:viso}. 
The short dashed lines show shells with  
$f_{\rm sh}=0.1$ and $f_{\rm sh}=1.0$ models (eq.~\ref{fsh}), but including a 
surrounding constant density medium of $n_{\rm ext}=10^{-4}$\,cm$^{-3}$ as motivated by
the hot halo models of \cite{maller_bullock}.  Both models attain high velocities because
$n_{\rm ext}<n_{\rm ext,\,crit}$, but then decelerate as they accumulate more mass, 
effectively stopping after $10^8$\,yr of evolution.
In a isothermal potential, these shells fall back again, particularly since the 
radiation pressure driving is not likely to be strong for timescales much larger $10^7-10^8$\,yr.
For the models shown, $L$ is held constant in time. 

The dotted line in Figure \ref{figure:viso} shows the evolution of the $f_{\rm sh}=1$ model
with a constant density external medium, but also with an isothermal sphere gas reservoir
of the form in equation (\ref{isodensity}).  For $f_g=0.01$, the maximum velocity of the shell
 is significantly decreased, as predicted from equation (\ref{fgcrit}) and for $f_g=0.1$ the velocity of the shell only reaches  $\sim300$\,km s$^{-1}$.

Note that in integrating the evolution of these shells we have had to make an assumption
about the dust content per unit mass of the swept up material, adjusting the UV and IR 
opacities accordingly. We have adopted a simple parameterization by assuming that the
dust-to-gas ratio of the swept up material is a constant, normalized to the Milky Way 
value: 
\beq
\xi=f_{\rm dg,\,swept}/f_{\rm dg,\,MW}.
\label{xi}
\eeq
We adjust the UV and IR opacities by averaging $\xi$
over the total mass of the shell, as it sweeps up ambient gas: 
\beq
\langle\xi\rangle=\int \xi(M_{\rm sh})\,dM_{\rm sh}/\int\,dM_{\rm sh}.    
\eeq
All the models in black in Figure \ref{figure:viso} that interact with an ambient medium
assume  $\xi=0.1$, but the results are not qualitatively different if we assume the swept up medium is 
completely dust-less, $\xi=0$.  As an example, for $f_{\rm sh}=1$, $f_g=0.1$, and $n_{\rm ext}=10^{-4}$ cm$^{-3}$(long dashed lines) we show models for $\xi=0$ and  $\xi=1$ in gray.   The other models shown are not as strongly effected by this change in $\xi$.

In addition to the calculations presented in Figure \ref{figure:viso}, we have done a number of tests
with a more realistic NFW dark matter potential.  Because the density profile
is steeper on large scales, the shells launched in the NFW potential generally 
attain higher asymptotic velocity than those launched in a singular isothermal sphere, all
else being equal.  However, it is clear from  Figure \ref{figure:viso} that the dynamics of 
shells is dominated by the large-scale gas distribution, and not the large scale potential.  The
maximum velocity of a shell, its velocity profile, and its long-term evolution depend sensitively
on both the radial dependence of the ambient density profile and its normalization.  For this 
reason, we have opted to focus on the simpler isothermal case, for which some analytic 
estimates are easily made.

\begin{figure*}
\centerline{\includegraphics[width=8.5cm]{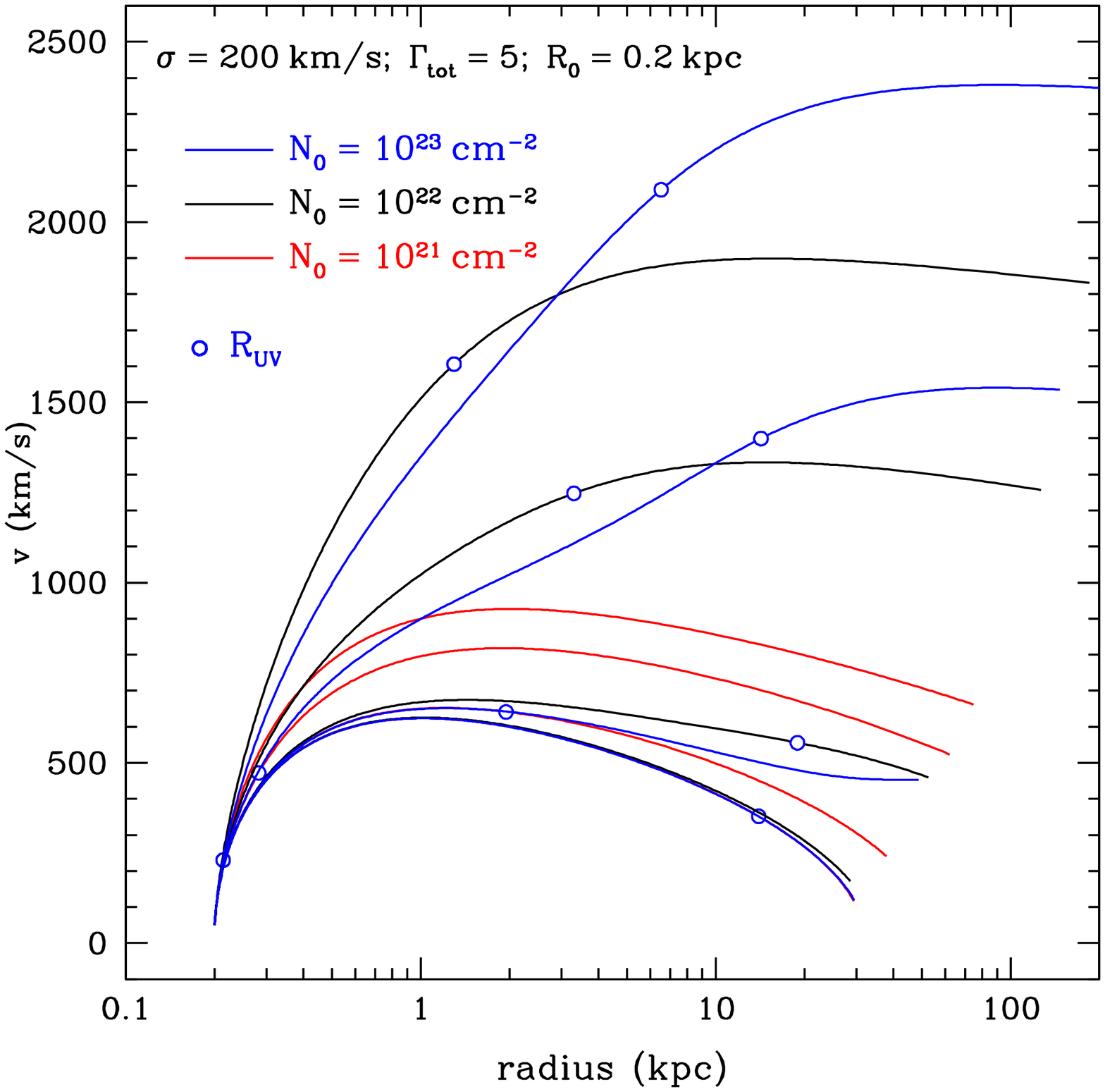}
\includegraphics[width=8.5cm]{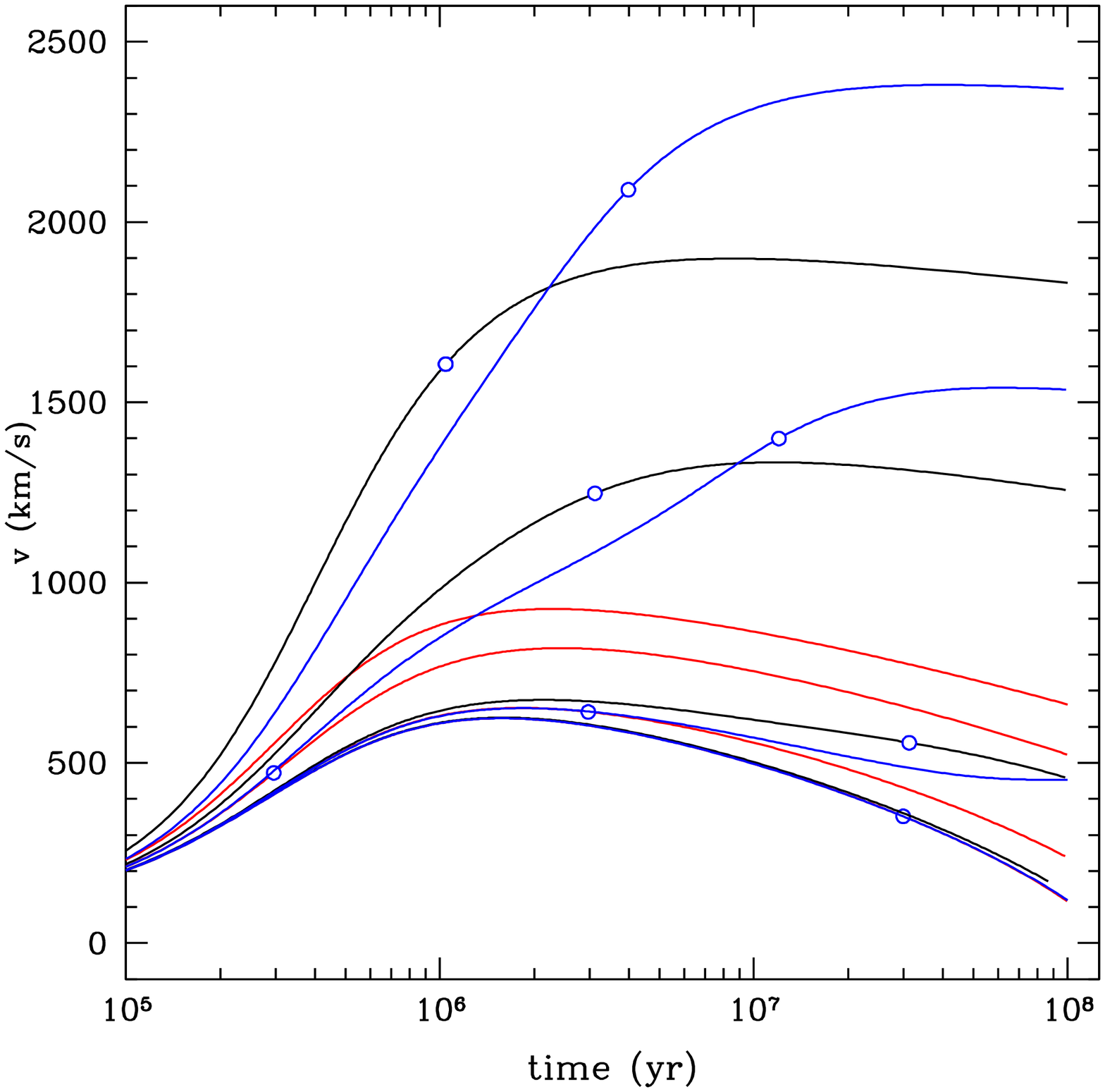}}
\caption{Velocity as a function of radius (left panel) and time (right panel) 
for dusty clouds with column densities $N_0=10^{21}$ (red lines), $10^{22}$ (black lines), and $10^{23}$\,cm$^{-2}$ (blue lines), expanding into vacuum with internal sound speed $c_s=0.1$, 1, 10, and 30\,km s$^{-1}$ (lowest to highest for each $N_0$) in an isothermal potential with $\sigma=200$\,km s$^{-1}$.  All clouds start from a launch radius of $R_0=0.2$\,kpc, have initial cloud radius of $10$\,pc, outward velocity of $50$\,km/s, and $\Gamma_{\rm tot}=5$ (eqs.~\ref{momcloud} \& \ref{leddcloud}). At fixed $\Gamma_{\rm tot}$, the total luminosities are different for each initial cloud column density.  They  are $L\simeq2.2\times10^{13}$, $3.8\times10^{12}$, and $5.0\times10^{11}$\,L$_\odot$ for $N_0=10^{23}$, $10^{22}$, and $10^{21}$\,cm$^{-2}$, respectively. The blue circles denote the radius $R_{\rm UV}$ where the UV optical depth of the cloud drops to unity.}
\label{figure:clouds}
\end{figure*}

\subsection{Clouds}
\label{section:clouds}

In the results presented above and below, we focus on the limit of a geometrically thin shell that subtends $4\pi$ for simplicity.   For clouds that cover only a small fraction of the solid angle from the source, the dynamics change quantitatively and qualitatively because a new timescale enters the problem: the cloud expansion time $t_{\rm exp}=(d\ln R_c/dt)^{-1}$, where $R_c$ is the cloud radius.  In the shell limit, the timescale for the shell to change its optical depth $\tau$ is directly coupled to the dynamical timescale because  $\tau\propto r^{-2}$, and this determines $R_{\rm UV}$, which in turn determines the asymptotic velocity.  For a cloud, $\tau\propto R_c^{-2}$ and the radius at which it becomes optically-thin to the UV is instead governed by the cloud's internal sound speed $c_s$ and the amount by which the cloud is over- or under-pressured with respect to the background medium. Cloud acceleration by radiation pressure on dust has been discussed in models of galactic winds by \cite{mqt,mmt}.

The set of equations governing the dynamics of a radiation pressure accelerated cloud expanding into vacuum (no background medium) is 
\beqa
\frac{d}{dt}\left(M_c\,v_c\right)&=&-\frac{GMM_c}{r^2}\nonumber \\
&+&\left(1+\min[1,\tau_{\rm IR}]-e^{-\tau_{\rm UV}}\right)
\frac{L}{c}\,\left(\frac{A_c}{4\pi r^2}\right)
\label{momcloud}
\eeqa
and 
\beq
\frac{d R_c}{dt} = c_s,
\label{radiuscloud}
\eeq
where we assume a geometrically thin ``pancake'' geometry for the cloud, $M_c$ is the cloud mass, $A_c=\pi R_c^2$, $\tau_{\rm UV,\,IR}=M_c\kappa_{\rm UV,\,IR}/A_c$, and we limit the $\tau_{\rm IR}$ force terms such that even if $\tau_{\rm IR}\gg1$ the cloud does not feel an enhanced force.  In reality, this term will depend on the distribution of clouds in the system since the optical depth for the cloud ensemble may be much larger or smaller than unity, and may be much different than the individual cloud optical depth.  

Setting the momentum equation equal to zero, we derive the generalized Eddington limit for clouds:
\beq
L_{\rm Edd}=\frac{4\pi G M c}{\left(1+\min[1,\tau_{\rm IR}]-e^{-\tau_{\rm UV}}\right)\,(A_c/M_c)}.
\label{leddcloud}
\eeq
Note that the denominator varies from $\simeq 2A_c/M_c$ (the single-scattering limit) as the cloud column density goes to infinity, to $\kappa_{\rm UV}$ ($L_{\rm Edd}=4\pi GM c/\kappa_{\rm UV}$) as the cloud column goes to zero.   The former limit is different than the shell geometry because of the $\tau_{\rm IR}$ term.  In the shell case, the Eddington luminosity approaches a constant ($L_{\rm Edd}=4\pi GM c/\kappa_{\rm IR}$), whereas in the cloud case $L_{\rm Edd}$ continues to increase with increasing cloud column since we impose the condition that $\tau_{\rm IR}$ cannot be larger than unity.

There are additional important similarities and differences between the dynamics of shells and clouds we wish to highlight.  As in the case of shells, we see that if $L\gtrsim L_{\rm Edd}$ and if the cloud starts optically-thick to the incoming radiation and expands as a function of radius (e.g., $R_c\sim t c_s$), because the cloud column decreases as a function of time, the cloud becomes increasingly super-Eddington as $r$ increases, like a sail continuously unfurling in the wind.   This has two consequences.  First, because a small perturbation in the radial direction causes the cloud to expand and the Eddington luminosity to drop, clouds are dynamically unstable to expulsion in the single-scattering limit, as discussed for shells in Section \ref{section:point}.  Second, the asymptotic velocity can be very large, as in the case of shells treated throughout this work.  The simplest way to see this is to note that {\it if} $R_c\propto r$, then the ratio $A_c/r^2=\,$constant on the right hand side of equation (\ref{momcloud}) and the solution for the cloud dynamics reduces to the shell case worked out in Section \ref{section:point}. However, in general $R_c$ is not a power-law in $r$ and this prevents a simple analytic treatment.

In the opposite limit where the cloud does not expand or simply has a very long expansion timescale relative to the dynamical timescale $r/v$, then $A_c/M_c$, $\tau_{\rm UV}$, and $\tau_{\rm IR}$ are effectively constant, and the solid angle subtended by the cloud decreases as $r^{-2}$ as the cloud is driven outward.  The asymptotic velocity of the cloud in the point-mass limit is then simply $v_\infty=v_{\rm esc}(R_0)\left(\Gamma-1\right)^{1/2}$, where $\Gamma$ is the initial value of $ L/L_{\rm Edd}$ (in eq.~\ref{leddcloud}).

These limits are illustrated in Figure \ref{figure:clouds}.   We show velocity as a function of radius and time for radiation pressure accelerated clouds with $\Gamma_{\rm tot}=L/L_{\rm Edd}=5$ (eq.~\ref{leddcloud}), with initial column densities of $N_0=10^{21}$ (red), $10^{22}$ (black), and $10^{23}$\,cm$^{-2}$ (blue), expanding into vacuum with a range of internal sound speeds from $c_s=0.1-30$\,km/s (lowest to highest) in an isothermal potential.  All clouds start with the same radius of $10$\,pc.  As expected, for small $c_s$, all clouds have approximately the same velocity profile, regardless of their initial column density, and their maximum velocity is $\sim 2\sigma(L/L_{\rm Edd}-1)^{1/2}$ (with a modification for the extended gravitational potential).  However, for higher cloud expansion speeds, the velocity evolution changes significantly, with much higher velocities attained, and depends on the initial column density of the cloud.  

The fastest expanding, lowest column density cloud (highest red line) reaches only $\sim800$\,km s$^{-1}$ because the cloud rapidly becomes optically-thin to the UV radiation, because it does not have a large value of $L/L_{\rm Edd}$ at $R_{\rm UV}$, and because it does not undergo a long-lived phase in the single-scattering limit. In contrast, the most rapidly expanding higher column clouds reach much higher velocities (highest blue and black lines).  The radius (or time) at which these clouds become optically-thin to the UV is indicated by the blue circled dots, as in earlier Figures, and shows both that the extended single-scattering phase is important and the value of $L/L_{\rm Edd}$ at $R_{\rm UV}$. 

All else being equal, clouds that start with cloud radius smaller than the 10\,pc value used in Figure \ref{figure:clouds} reach higher velocities because the cloud expansion timescale is shorter.  For the $N_0=10^{22}$\,cm$^{-2}$ clouds, for example, if the cloud starts with a radius of 1\,pc and $c_s=10$\,km s$^{-1}$, the cloud reaches an outward velocity $\sim2500$\,km s$^{-1}$ instead of the $\sim1300$\,km s$^{-1}$ shown in Figure \ref{figure:clouds}.  In this regime of rapid expansion the cloud becomes optically-thin to the UV radiation on small scales and the effective Eddington ratio at $R_{\rm UV}$ is large, leading to a higher value of the asymptotic velocity.  In this example, for a cloud that is optically-thin to the UV near $R_0$, the effective Eddington ratio would be $\sim80$ instead of $5$.  The rapid expansion thus allows for high acceleration in the super-Eddington flux from the source. 

Note that we have assumed that there is no surrounding medium and that the clouds are not destroyed by the Kelvin-Helmholz instability or evaporation (see, e.g., \citealt{cooper09}).  We have further assumed a single constant internal sound speed and that the UV light from the central source is unobscured.  In fact, we expect clouds with a range of column densities to be driven out of the system with a spread of asymptotic velocities (see \citealt{thompson_krumholz}) and for the ensemble of clouds to partially obscure the central source, thereby limiting the  acceleration.  In addition, there is ample evidence for a hot gas component in galactic winds that the cold clouds will sweep up and interact with, potentially leading to their destruction and a different radial evolution than indicated by equation (\ref{radiuscloud}) (\citealt{strickland_stevens,strickland_heckman,zhang14,zhang15}).  These effects require further study.

\section{Discussion}
\label{section:discussion}

Below, we provide brief discussions of the total momentum in radiation pressure driven flows, the applicability of our results to observations of fast outflows observed in emission and absorption in local and high-redshift galaxies and AGN, prescriptions for outflows in large-scale cosmological simulations, and the implications of our results for star cluster disruption and massive star eruptions.  We focus on shells throughout instead of clouds since the dynamics of the latter require a more complete understanding of how the dynamics change when an ensemble of clouds of different column densities, radii, and velocities obscure the central source.  This important problem will be the focus of a future investigation.

\subsection{The Asymptotic Momentum of Shells}
\label{section:momentum}

A key diagnostic of observed outflows in galaxies and AGN is the
momentum ratio 
\beq
\zeta=\frac{\dot{M}v}{(L/c)}=\frac{M_{\rm sh}v}{(L/c)(r/v)},
\label{momratio}
\eeq
where the first equality is applicable to a continuous wind with mass loss rate $\dot{M}$,
and the second equality is applicable to a shell.  These two definitions
are equivalent since, typically, one measures
the column density in blue-shifted absorption features 
(such as the resonance lines of Fe, Mg, and Na) to infer
\beq
\dot{M}=4\pi r N m_p v  \,\,\,\,{\rm or}\,\,\,\, 
M_{\rm sh}=4\pi r^2 N m_p,
\eeq
where $N$ is the column density of gas, so that 
\beq
\zeta = \frac{4\pi r N m_p v}{(L/c)}
\eeq
in either case.  In general, one must assume a relative abundance of the tracer 
(e.g., Na or Mg) with respect to total gas, which usually involves
an uncertain ionization correction (e.g., \citealt{murray07}).  

A primary observational difference between a continuous wind and a shell would of course 
be in the absorption line profile, which for a perfect geometrically-thin single shell would be a 
delta-function in velocity along the line of sight toward a point source of radiation.  However,
for a shell in proximity to a finite-sized  source (a star or galaxy), the 
observed absorption line would be broadened
geometrically by the projection of the moving shell onto the source.

The momentum ratio $\zeta$ is important because for $\zeta\gg 1$, either (1)
the shell or wind had very high effective $\tau_{\rm IR}$ (e.g., \citealt{mqt10}), or 
(2) the shell was initially energy-driven, as in the early evolution of a supernova 
remnant.  Since the radial scale 
of the absorbing material is in general not known, absorption
line studies determine $\zeta$ with significant uncertainties.  
Emission-line studies with molecular emission, [CII], or in the optical/UV (e.g., H$\alpha$, [NII]) 
provide a complementary view of winds, and in some cases find
evidence for $\zeta>1$ (e.g., \citealt{cicone}; \citealt{genzel11}; Section \ref{section:outflows}).

\begin{figure*}
\centerline{\includegraphics[width=8.5cm]{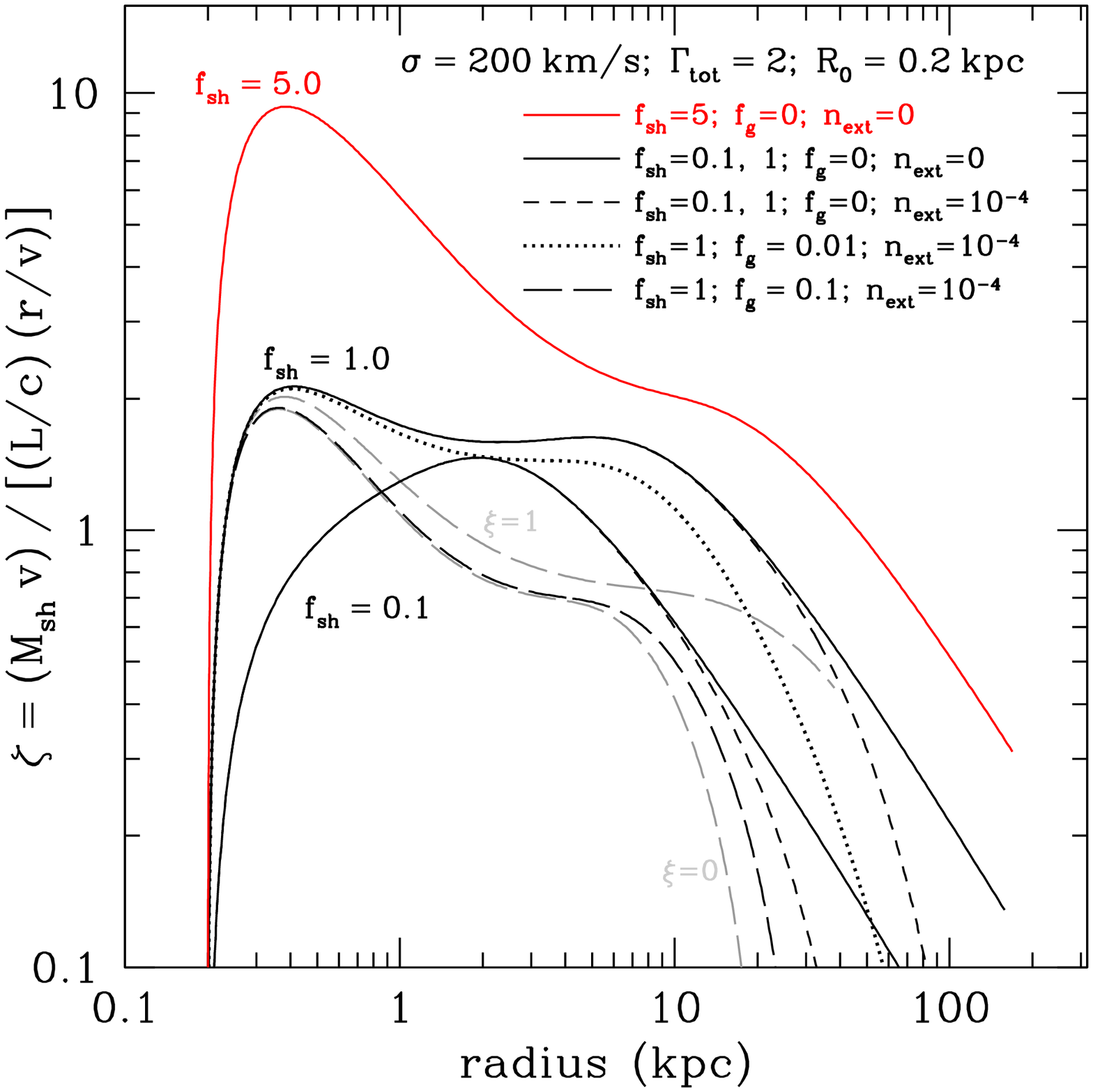}
\includegraphics[width=8.5cm]{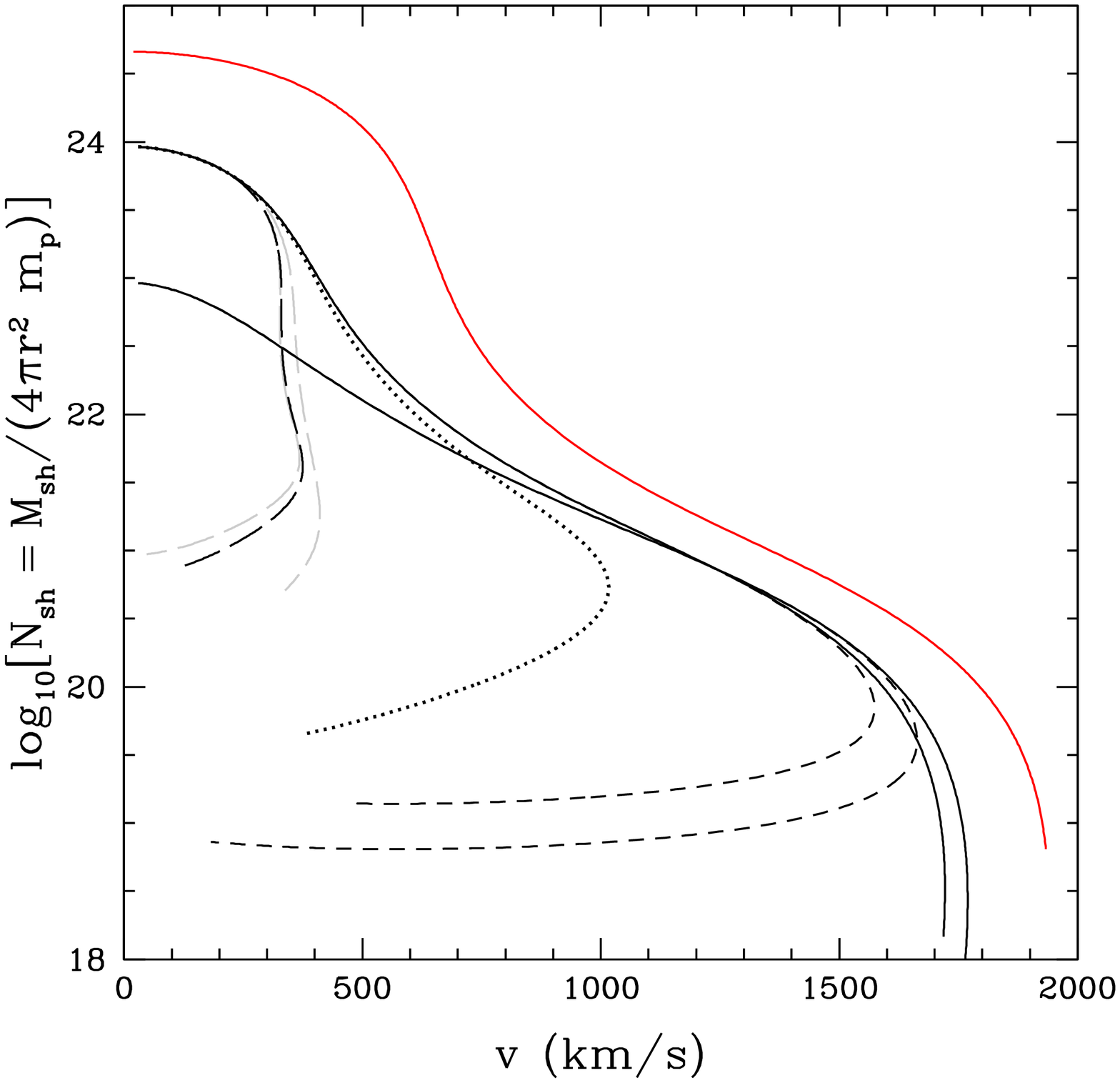}}
\caption{(Left): Momentum ratio $\zeta$ (eq.~\ref{momratio}) as a function of radius for
the shells shown in Figure \ref{figure:viso}, but with an additional high-$\zeta$ model 
for illustration with $f_{\rm sh}=5$ (red line), which has $\tau_{\rm IR}(R_0)\simeq40$.
(Right):  Total hydrogen column density of shells as a function of velocity.
Models and linestyles are the same as in the left panel and Figure \ref{figure:viso}.}
\label{figure:ratio}
\end{figure*}

The momentum ratio in the case of a single shell of fixed mass, 
observed at a radius $r\gtrsim R_{\rm UV}$, can be approximated by 
\beq
\zeta=\frac{M_{\rm sh}v_{\rm UV}^2}{rL/c}\simeq2f(\tau_0)\,\left(\frac{\Gamma_{\rm SS}}{\Gamma_{\rm tot}}\right)\left(\frac{R_{\rm UV}}{r}\right),
\label{ratiouv}
\eeq
where $f(\tau_0)=(1+\tau_{\rm IR}-e^{-\tau_{\rm UV}})$ evaluated at the launch radius $R_0$.  
For $\tau_{\rm UV}>1$, but $\tau_{\rm IR}<1$, $\Gamma_{\rm SS}\simeq\Gamma_{\rm tot}$ and $f(\tau_0)\sim1$, implying 
that $\zeta$ should always be of order unity or smaller, since for $r> R_{\rm UV}$, $\zeta$ rapidly 
decreases.  This behavior can be seen in both the $f_{\rm sh}=0.1$ and the $f_{\rm sh}=1$ models
shown by the solid lines in Figure \ref{figure:ratio} (left), which shows $\zeta(r)$.  All models 
have $\zeta\sim0.1$ on $10-100$\,kpc scales.  Note that even though the shells accelerate
beyond $R_{\rm UV}$, the measured momentum decreases, because in this regime $v^2(r)/r$ decreases.
Models that sweep up mass decelerate, and $\zeta(r)$ decreases more rapidly.  In the 
single-scattering limit of $\tau_{\rm UV}>1$, but $\tau_{\rm IR}<1$ at $R_0$, one would
simply estimate $v^2_{\rm UV}\sim R_{\rm UV} L/M_{\rm sh} c$ by dimensional analysis
of the momentum equation, and then substituting into equation (\ref{ratiouv}), one again finds
that $\zeta(R_{\rm UV})\simeq1$.

Since the two key observables for characterizing shell-like outflows are velocity and column
density, and since these enter the calculation of $\zeta$ directly, in the right panel of 
Figure \ref{figure:ratio} we show $M_{\rm sh}/4\pi r^2 m_p$ versus $v$ for the same models as 
in the left panel, and in both panels of Figure \ref{figure:viso}.  Note that, as expected, the 
momentum ratio reaches $\sim1$ for the shells dominated by the single-scattering term
and then decreases on larger scales, where the shell spends the most time.  If radiation 
pressure driven shells are a good model for fast outflows, $\zeta<1$ should be observed in some
systems. More discussion is provided in 
Section \ref{section:outflows}.

If the effective IR optical depth of the shell is much larger than unity at the launch point (see Section \ref{section:uncertain} below
for caveats), then $f(\tau_0)\sim\tau_{\rm IR}(R_0)$, $\Gamma_{\rm tot}\sim\Gamma_{\rm IR}$ and $\zeta(r)$ can be significantly increased.  For the 
$f_{\rm sh}=1$ example in Figures \ref{figure:viso} and \ref{figure:ratio}, $\tau_{\rm IR}(R_0)\simeq8$ and $\zeta(r)$
peaks at $\simeq2$ on the scale of a few times $R_0$.  For higher $\tau_{\rm IR}(R_0)$ one finds higher 
values of $\zeta$ on the scale of $R_{\rm IR}$.  In particular, when $\tau_{\rm IR}(R_0)\gg1$ and $\Gamma_{\rm IR}\gg1$, one finds that at $R_{\rm IR}$
\beqa
\zeta(R_{\rm IR})&=&\frac{M_{\rm sh}v_{\rm IR}^2}{R_{\rm IR}L/c}
\simeq2\frac{R_{\rm IR}}{R_0} \nonumber \\
&\simeq&5.8\,\,\kappa^{1/2}_{\rm IR,\,0.7}\,M_{\rm sh,\,9}^{1/2}\,R_{\rm 0,\,0.1kpc}^{-1}.
\eeqa
Note that while $\zeta(R_{\rm IR})$ is not proportional to $\tau_{\rm IR}(R_0)$, the maximum value of $\zeta$ (see Fig.~\ref{figure:ratio}), which  occurs on scales smaller than $R_{\rm IR}$, is proportional to $\tau_{\rm IR}(R_0)$ (for fixed $\Gamma_{\rm IR}\gg1$).  The red solid lines in Figure \ref{figure:ratio} show $\zeta(r)$ and $N(v)$ for a model with $f_{\rm sh}=5$ so that $\tau_{\rm IR}(R_0)\simeq40$.  The peak in $\zeta(r)$ occurs at a few times $R_0$ and is about 4 times lower than $\tau_{\rm IR}(R_0)$.  See Section \ref{section:uncertain} for a discussion of some of the uncertainties associated with high-$\tau_{\rm IR}$ solutions in the context of radiation pressure driven shells.

\begin{figure*}
\centerline{\includegraphics[width=8.5cm]{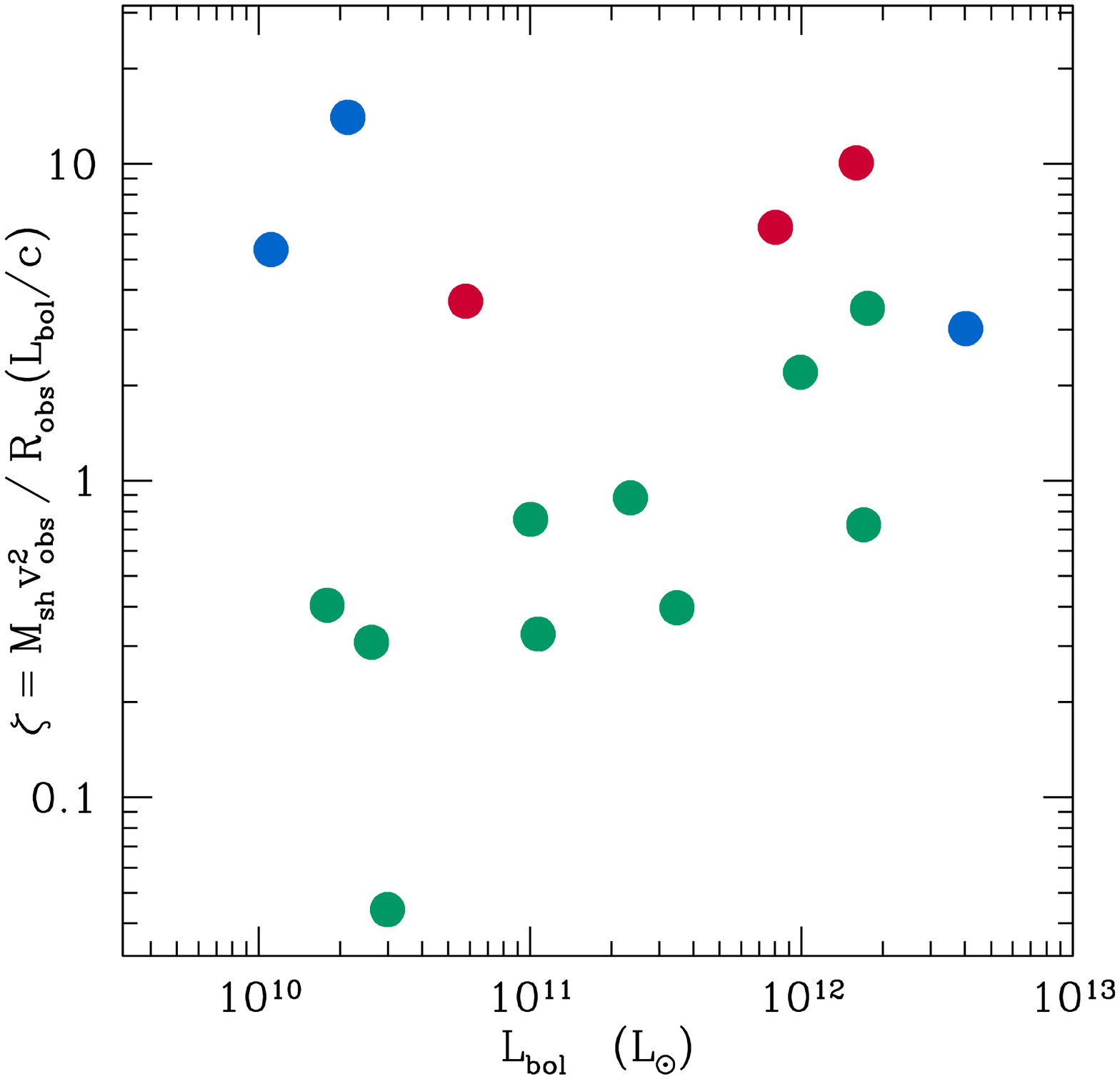}
\includegraphics[width=8.5cm]{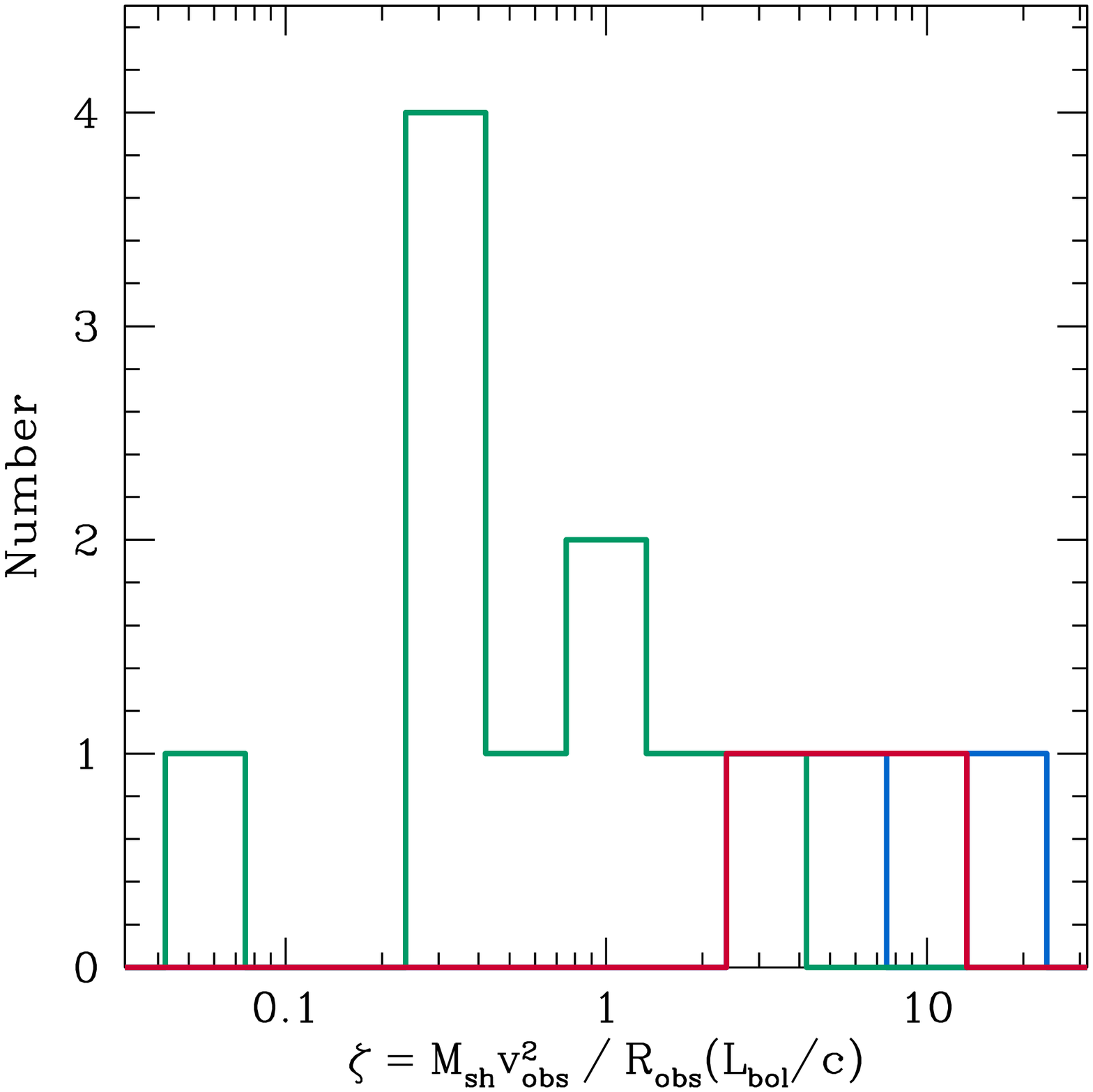}}
\caption{(Left): Observed momentum ratio from a sample of AGN and starburst galaxies 
as a function of bolometric luminosity \citep{cicone}.  Red, blue, and green points show systems 
where the central AGN luminosity $L_{\rm AGN}/L_{\rm bol}>0.5$, $0.2\leq L_{\rm AGN}/L_{\rm bol}\leq 0.5$, 
and $L_{\rm AGN}/L_{\rm bol}<0.2$, respectively. (Right):  Histogram of momentum ratios from the same sample.  
A CO-to-H$_2$ conversion factor of 0.8 has been used for calculating $M_{\rm sh}$.}
\label{figure:c}
\end{figure*}

\subsection{Fast Outflows from Rapidly Star-Forming Galaxies,  Starbursts, \& Post-Starbursts}
\label{section:outflows}

{\bf Fast Outflows in Emission:}
\cite{cicone} have recently presented data on a collection of AGN and star-formation dominated
systems with outflows seen in molecular emission.  We show $\zeta$ as a function of $L_{\rm bol}$
and a histogram of $\zeta$ for their sample in the panels of Figure \ref{figure:c}.  Red, blue, and green denote AGN fractions
of $L_{\rm AGN}/L_{\rm bol}>0.5$, $0.2\leq L_{\rm AGN}/L_{\rm bol}\leq0.5$, and $L_{\rm AGN}/L_{\rm bol}<0.2$, respectively, for the systems
with outflows detected at high significance.\footnote{By assuming a uniform medium,
\cite{cicone} overestimate $\zeta$ by a factor of 3, which  we have corrected in Figure \ref{figure:c}.
We also adopt the lowest values of the total inferred outflowing gas mass in their tables.  In some
cases $\zeta$ could be larger by a factor of $\sim3$.}

If interpreted as shells accelerated by radiation pressure on dust
we find that the systems with $\zeta\lesssim2-3$ are readily explained, given their bolometric 
luminosities.  For example, Mrk 273 and 231, have 
$L_{\rm AGN}/L_{\rm bol}\simeq0.08$ and $\simeq0.3$, 
$L_{\rm bol}\simeq1.7\times10^{12}$ and $4\times10^{12}$\,L$_\odot$, and
total outflow mass of $\sim10^{8.2}$ and $\sim10^{8.4}$\,M$_\odot$, respectively, on scales of $\simeq0.6$\,kpc,
with average velocities of $\simeq600-700$\,km s$^{-1}$.  These parameters 
are all in the range expected for relatively low-mass shells accelerated by radiation pressure, with
dynamics similar to the $f_{\rm sh}=0.1$ model shown in Figures \ref{figure:viso} and \ref{figure:ratio}.

Systems with high values of $\zeta$ in the \cite{cicone} compilation likely 
require large initial values of $\tau_{\rm IR}(R_0)$ (see red line in Figure \ref{figure:ratio}; 
Sections \ref{section:momentum} and \ref{section:uncertain}).   Another possibility is 
that a fairly rapid decrease in the AGN or starburst luminosity could imply large $\zeta$
even though the dynamics is consistent with radiation pressure acceleration
in the single-scattering limit.   Future explorations
could employ the shell models described here and/or continuous wind models to constrain
$R_0$ and the source of driving with the data on $\zeta$.  A careful comparison with 
the observed dynamics could be used to quantitatively test the radiation pressure driven 
shell picture discussed here. 

{\bf Fast Outflows in Absorption:}
\cite{tremonti} and \cite{diamond} report the discovery of fast $\gtrsim1000$\,km s$^{-1}$
outflows in starburst and post-starburst galaxies.  In particular, \cite{diamond} show that 
the system J0905+5759 has strongly blue-shifted Mg II absorption with a shell-like
velocity profile centered at $-2470$\,km s$^{-1}$ (ranging from $\simeq-3000$ to $-2200$\,km s$^{-1}$), covering the entire galaxy.  The effective
radius of the galaxy is $\simeq100$\,pc, with stellar mass of $10^{10.7}$\,M$_\odot$, velocity 
dispersion of $\simeq250$\,km s$^{-1}$, and total IR luminosity of $10^{12.6}$\,L$_\odot$
(Diamond-Stanic, private communication; see also \citealt{sell,geach}).

Repeating the calculation shown in the left panel of Figure \ref{figure:viso}, but for $\sigma=250$\,km
s$^{-1}$, $R_0=100$\,pc, and $\Gamma_{\rm tot}=2.5$ ($L_{\rm bol}\simeq1.1\times10^{13}$\,L$_\odot$) we find $v_{\rm UV}\simeq1900$\,km s$^{-1}$ and $v_\infty\simeq2500$\,km s$^{-1}$ for $f_{\rm sh}=0.05$ ($M_{\rm sh}\simeq1.5\times10^8$\,M$_\odot$).  If interpreted as a shell accelerated by the
continuum radiation pressure on dust grains, the shell would have a radial scale of $\gtrsim1.6$\,kpc,
which is consistent with the observed blue continuum.

We note that although the velocity and shell mass calculated are roughly consistent with the 
observational constraints,  the required bolometric luminosity of the system is 
somewhat high compared to the observations.  
However, there are many factors that complicate straightforward modeling 
of this system.  In particular, a realistic model for the luminosity of the system as a function of 
time and the large-scale gravitational potential change the velocity profile quantitatively.  More 
importantly, though, as highlighted in Figure \ref{figure:viso}, the surrounding medium swept 
up by the shell can qualitatively effect its dynamics.  In this context, it is again worth noting
that the Eddington ratio for dusty shells is linearly dependent on the gas-to-dust-ratio: larger 
dust content per gram of gas lowers the critical luminosity for shell expulsion.  Finally, the absorption 
line profile for this system has significant width in velocity \citep{diamond}, which may indicate that the 
cloud picture discussed in Section \ref{section:clouds} is more applicable than the case of a simple shell.
Future work should model the system with an ensemble of absorbing clouds accelerated to a
range of velocities (see Fig.~\ref{figure:clouds}) to see if the observations
can be reproduced.

{\bf The Local Starburst M82:}
M82 has an extensively studied outflow, 
with evidence for $\sim200-600$\,km s$^{-1}$ line-emitting gas 
and dust on kpc scales, a molecular outflow on small scales,  
and hot nuclear X-ray emission (e.g., \citealt{shopbell,walter,strickland_heckman}).  

Using the dust-scattered UV emission from the central starburst \citep{hoopes}, 
\cite{coker} calculated the Eddington ratio for dusty gas on $0.5-5$\,kpc scales.  
Using the large-scale rotation curve from \cite{greco}, \cite{coker} found
that although significantly more UV escapes the starburst along the minor axis than along
our line of sight, the Eddington ratio is still much less than unity on large scales.
These results imply that a shell-like radiation pressure-driven outflow cannot account
for the dusty gas currently seen on kpc scales.  Moreover,  this result
highlights the fact that the additional large-scale
acceleration that  produces the high velocities discussed here may 
not be generic, but may only occur in special circumstances or geometries.
One mitigating factor is that \cite{forster_schreiber} find that 
the bolometric luminosity of the M82 starburst was a factor of $\sim4$ larger 6\,Myr ago, 
indicating that the system may have potentially approached the single-scattering Eddington 
limit on scales within the starburst\footnote{
$\Gamma_{\rm SS}\sim1$ for 
$M(r<300\,{\rm pc})=10^9$\,M$_\odot$, $M_{\rm sh}=10^8$\,M$_\odot$, 
$R=300$\,pc and $L=2\times10^{11}$\,L$_\odot$ (see eq.~\ref{gammass}).
},  and would have certainly exceeded $\Gamma_{\rm UV}=1$
(eq.~\ref{gammauv}).

\subsection{Broad Absorption Line Quasars \& Ultra-Fast Outflows}

Some BAL quasars have detached potentially shell-like absorption profiles with blueshifted 
velocities of $\simeq4000-5000$\,km s$^{-1}$.  Some have velocities over
$20,000$\,km s$^{-1}$ (e.g., \citealt{pounds03,tombesi10,gupta13}).  

Taking $L_{\rm AGN}=10^{47}$\,ergs s$^{-1}$ and $R_0=R_{\rm sub}\simeq2$\,pc
and $M_{\rm sh}=10^5$, $10^6$, and $10^7$\,M$_\odot$, we find 
$v_\infty\simeq2.8\times10^4$, $1.7\times10^4$ and $1.1\times10^4$\,km s$^{-1}$, respectively,
for Milky Way gas-to-dust ratio.  Such high
velocity outflows might emerge along the line of sight to Type-I quasars in a very short
timescale $\ll 10^6$\,yr (see eq.~\ref{tacc}) and become optically-thin to the UV emission from the quasar
on scales $<0.1-1$\,kpc (see eq.~\ref{ruv}).  These types of dusty shells are related to the suggestion
by \cite{scoville_norman} that dusty material near quasars could be accelerated to $\sim0.1$\,c.

\subsection{Previous Work \& Prescriptions for Cosmological Simulations}

\cite{mqt} discussed galactic winds driven by the combined momentum
input of radiation pressure on dust and supernovae and wrote down the 
momentum equation for a shell in an isothermal potential, as in equation
(\ref{genmom2}).  The primary difference between their results and our
work here is in the assumed distribution of mass swept up by the outflowing shell.	
In particular, they assumed an isothermal mass distribution
for the gas so that $M_{\rm sh}\propto r$ (as in eq.~\ref{isodensity}).  In this case, in the single-scattering
limit,
\beq
v\frac{dv}{dr}=-\frac{\sigma^2}{r}+\frac{L}{c}\frac{G}{2\sigma^2 f_g r}
\label{isogas}
\eeq
where $f_g=M_{\rm sh}/M$ is a constant at each radius.  Note that for 
constant $f_g$, both the gravitational acceleration and the radiation pressure
acceleration have the same radial dependence.  This is why the fairly large enhancement
in the asymptotic velocity of a shell launched by radiation pressure we emphasize
here was not noted in that work.  The long dashed line in the panels of
Figure \ref{figure:viso} assumes an isothermal sphere for the surrounding gas 
distribution and shows a shell model that closely tracks the expectation from equation (\ref{isogas}) 
and  \cite{mqt}.

Several studies of the enrichment of the intergalactic medium have used 
the so-called "momentum scalings"  (radiation pressure or supernovae) 
based on the work of \cite{mqt}  (e.g., \citealt{oppenheimer_dave06,oppenheimer_dave08}).
Typically, 	these are that 
\beq
v_\infty\simeq{\rm few}\times\sigma
\label{vsigma}
\eeq
and that\footnote{This follows
from equating $L/c\simeq\epsilon\,{\rm SFR}\,c\sim\dot{M} v_\infty$ 
for a star-forming galaxy in the single-scattering limit \citep{mqt}.}
\beq
\dot{M}/{\rm SFR}\propto \sigma^{-1}
\label{mdotsfr}
\eeq
These same scalings and variants have also been tested in models of the
mass-metallicity relation by \cite{peeples_shankar}.  

Our work suggests prescriptions that would more accurately capture
the physics of radiation pressure driven shells, as opposed to continuous winds.  
The simplest is that the ``few'' in equation (\ref{vsigma}) could in some cases be
as large as $\sim10$, depending on the geometry of the outflow and the surrounding medium,
a potentially important factor that should be taken into account and explored in more detail.
Second, the Eddington luminosity for a shell
is given simply by equation (\ref{leddiso}), which can be thought of as three different
Eddington luminosities in three different regimes.  If the shell is (1) optically-thick to the
IR, (2) optically-thin to the IR, but optically-thick to the UV, or (3) optically-thin to the UV, the
Eddington luminosity is given by equations (\ref{gammair}), (\ref{gammass}), and 
(\ref{gammauv}), respectively, depending on the surface density of the 
shell.  If $L\geq L_{\rm Edd}$ in the appropriate limit, then the ISM is ejected.  Note
that in general $L_{\rm Edd}$ is dust-to-gas ratio dependent and thus metallicity dependent.

The dynamics of the shell and its interaction with the surrounding circumgalactic medium
could be calculated from equation (\ref{genmom}), either via a subgrid model in 
large-scale cosmological simulations, or explicitly in high-resolution simulations of 
individual galaxies.  Such a prescription for large-scale simulations
would differ qualitatively from what is currently
done in that a large fraction of the ISM would be ejected in single events, fallback would
be determined predominantly by the circumgalactic medium density profile, and the 
velocity of the material could approach $\sim10\times\sigma$ along lines of sight with 
little gas (see eq.~\ref{voutiso}).  In this picture, the ratio $\dot{M}/{\rm SFR}$ should
instead be thought of as the ratio of the total mass ejected to the total mass formed
between each star formation and ejection event, where the timescale between ejections
would be determined by the gas accretion rate from the IGM and from re-accreted 
(formerly ejected) gas.  Ejection episodes and fallback might precipitate radiative 
cooling of  the hot circumgalactic medium, as in the work of Fraternali \& Binney 2008
(see also Marasco et al.~2012; Fraternali et al.~2013).

\subsection{Star Cluster Disruption}

The estimates made here can also effect giant molecular cloud (GMC) disruption.
\cite{mqt10} discussed the acceleration of GMC gas by radiation pressure and other forces (see also \citealt{krumholz_matzner}), and \cite{mmt} provide a general description of launching these shells and clouds from star clusters to velocities high enough to escape the host galaxy and generate supershells.  In these works, analytic estimates for the critical star cluster stellar mass required to generate extra-planar gas was estimated, based on the assumption that $v_\infty$ for such a shell would be of order a few times the cluster escape velocity.  If the self-gravity of the shell dominates the total gravitational force, and if the central star cluster reaches the single-scattering Eddington limit ($\Gamma_{\rm SS}\simeq1$), then we have shown here that $v_\infty\simeq (2GM_{\rm sh}/R_0)^{1/2}(R_{\rm UV}/R_0)^{1/2}\simeq200\,{\rm km/s}\,M_{\rm sh,\,6}^{3/4}\kappa_{\rm UV,\,3}^{1/4}(5\,{\rm pc}/R_0)^{-1}$, implying that large-scale super-shells from GMC disruption could be driven high above the plane of a large galaxy by massive star clusters with total stellar mass significantly less than $10^6$\,M$_\odot$.  Faster asymptotic velocities are more easily obtained in a shell-like geometry.

Because shell formation during GMC disruption by radiation pressure may be generic \citep{yeh_matzner},
our results are important for diagnosing the dynamics of observed systems like 30 Doradus
\citep{lopez,pellegrini}.

\subsection{Massive Star Eruptions}
\label{section:massive}

Figure \ref{figure:v} shows that shells from massive star outbursts can be acccelerated to velocities much larger than the escape velocity at the dust formation radius.  Here, we briefly discuss applications to the outbursts of Luminous Blue Variables and the shells observed around the yellow hypergiants VY CMa and IRC 10420.

Eta Carinae's homunculus shows a number of different kinematic components.  The primary mass reservoir is $\sim10$\,M$_\odot$ with a velocity of $\simeq500$\,km s$^{-1}$ (Smith et al.~2003; Smith 2006).  There is also a faster component at $\sim1000-2000$\,km s$^{-1}$ and a much faster, but much less massive component moving at $\sim3000-6000$\,km s$^{-1}$ \citep{smith_fast}.  All are associated with the Great Eruption approximately 170 years ago in which Eta Car reached an estimated bolometric luminosity of $\sim2\times10^7$\,L$_\odot$ \citep{davidson_humphreys}.

In this context it is worth asking if radiation pressure on dust grains could have dominated the acceleration of any of the kinematic components.  Examining Figure \ref{figure:v} and equations (\ref{vx}) and (\ref{tacc}), this possibility appears unlikely for the more massive component because (1) the asymptotic velocity would only be $\sim250-350$\,km s$^{-1}$ (allowing for some uncertainty in $L$) and (2) the timescale to reach this velocity is too long, $\sim10^3$\,yr.  For the less massive high velocity components radiation pressure on dust might have had more of a role, depending on their mass.  Although not shown in Figure \ref{figure:v}, calculations with $L=10^7$\,L$_\odot$ and $M_{\rm sh}=10^{-2}$, $10^{-3}$\,M$_\odot$, and $10^{-4}$\,M$_\odot$ reach $v_\infty\simeq970$, $\simeq1650$, and $\simeq2650$\,km s$^{-1}$ on decade-to-year timescales (eq.~\ref{tacc}).\footnote{The photospheric temperature of Eta Carinae and the 
yellow hypergiants discussed here is lower than needed for significant UV emission.  For this reason,
$\kappa_{\rm UV}$ used throughout this work should be replaced by the flux-mean dust opacity 
for a $\sim5000-7000$\,K blackbody: e.g., $\kappa\simeq100-300$\,cm$^2$ g$^{-1}$.  
This lowers the expected asymptotic velocity according to the scalings in equations (\ref{vx}) and (\ref{vmax_thin}).
\label{foot:kappa}}

 Clearly only small amounts of mass can be accelerated to the requisite velocities on short timescales.   That these components might be dynamically accelerated by radiation pressure on dust is connected to treatments of line-driven photon-tired outflows in this context by \cite{owocki}.  It is worth noting that the maximum possible velocity for a dusty shell accelerated from the dust formation/destruction radius is 
\beqa
v_\infty&\simeq&\left(\frac{4L\,\kappa^2_{\rm UV}\sigma_{\rm SB}T_{\rm sub}^4}{\pi c^2}\right)^{1/4} \nonumber \\
&\sim&3500\,{\rm km\,s^{-1}}\,L_7^{1/4}\kappa_{\rm UV,\,3}^{1/2}T_{1500}
\label{vmax_thin}
\eeqa
for a shell that starts optically-thin to the incident UV radiation from the star and a nominal Milky Way gas-to-dust ratio
(compare with eq.~\ref{vx}).  Allowing for a factor of 2 higher luminosity and dust-to-gas ratio boosts $v_\infty$ to $\sim4200$\,km s$^{-1}$.

The much lower velocity dusty outflows of the yellow hypergiants VY CMa and IRC 10420 are also of interest.  The latter has 
$L\simeq5\times10^5$\,L$_\odot$ and a mass of $\sim10-20$\,M$_\odot$, with dusty shells observed at velocity $\sim40$\,km s$^{-1}$
on $\sim10^3$\,AU scales, with equivalent mass loss rates of $\sim10^{-4}-10^{-3}$\,M$_\odot$ yr$^{-1}$ 
\citep{oudmaijer,humphreys97,humphreys_10420,dinh}.  The observed velocity is large compared to 
the escape velocity at the dust formation radius, $\sim20$\,km s$^{-1}$ and the typical 
velocities of dusty AGB star winds.  The observed parameters for VY CMa are similar.  It has
a prominent dusty arc with velocity $\sim50$\,km s$^{-1}$ on $10^3$\,AU scales, but with
both slower ($\sim10$\,km$^{-1}$) and faster ($\sim100-200$\,km s$^{-1}$) material observed 
smaller and larger scales, respectively \citep{monnier,humphreys_vycma,muller_vycma}.

In accord with equations (\ref{vx}) and (\ref{tacc}), low-mass shells with $M_{\rm sh}=0.1$\,M$_\odot$ from a 
star with  $L=5\times10^5$\,L$_\odot$, and $M=15$\,M$_\odot$ can
be accelerated to $\sim50-75$\,km s$^{-1}$ on scales of $\sim10-100$\,AU, but reach an asymptotic velocity
of $\sim150$\,km s$^{-1}$ at $\sim10^4$\,AU.  Higher mass shells of $\sim1-3$\,M$_\odot$ reach only 
$v_\infty\simeq100$\,km s$^{-1}$.\footnote{Note that these models have shell masses only marginally 
below the critical photon tiring limit discussed in footnote \ref{tiring}.  For $M_{\rm sh}=3$\,M$_\odot$
and $L=5\times10^5$\,L$_\odot$, one finds that $Lt/(M_{\rm sh} v^2/2)\sim4$ on 100\,AU scales.  A lower $L$,
or a larger dust condensation temperature ($T_{\rm sub}=1500$\,K) than assumed here would put 
these models into the photon tiring limit \citep{owocki_gayley}.}  
Here, as throughout this paper, we assume that the shell subtends
$4\pi$ and that it sees the central source throughout its acceleration to $v_\infty$.  Both assumptions
should be called into question for these mass ejection episodes since the dusty nebulae are observed to 
be asymmetric and since multiple shells exist for both stars.  Even so, the large-scale acceleration 
that is the focus of this paper might be 
required to explain the $\sim100-200$\,km s$^{-1}$ material seen 
around VY CMa by \cite{humphreys_vycma} (their Fig.~13).

\subsection{Uncertainties \& Assumptions}
\label{section:uncertain}

{\bf Geometry \& Emergent Radiation Field:}  We assume
a simple geometry throughout most of this paper: a point source with a surrounding spherical shell,
or a cloud that subtends much less than $4\pi$ and expands at its internal sound speed.
In most contexts, a shell is unlikely to subtend $4\pi$ and may break up into discrete
clouds and the clouds may be subject to destruction by hydrodynamical processes if there is a background
medium.  Perhaps more importantly, in the case
of galactic-scale winds, the set of sources is not point-like, although the galaxy
may be represented by a distribution of actively star-forming and disrupting
star clusters. 

In addition, we have assumed that the UV continuum escapes from 
the source to large scales and that it is time-steady.  In the galaxy context, the central source may 
be obscured by the ISM of the galaxy, or by intervening shells of material, or many expanding
clouds at different velocities and column densities \citep{thompson_krumholz}.\footnote{
A related point is that throughout this work we have implicitly 
neglected the importance of ionizing photons, which carry 
approximately as much momentum as the UV continuum in the 
single-scattering limit.   If the medium
between the source and the shell was optically-thin to ionizing photons,
the dynamics of the shells would change since these photons
would couple to the shell until it reached a very low column density. }
Although we have shown that the acceleration time is short with respect to the
characteristic time for a stellar population to change luminosity
(eq.~\ref{tacc}), the time evolution could be important effect for 
the long-term dynamics in galactic potentials
($>10^7-10^8$\,yr).  Similar to the discussion presented in \cite{zhang_thompson}
it is worth noting that strong blue-shifted absorption can be observed 
in post-starburst galaxies because the timescale for material driven to 
$\sim100$\,kpc is longer than $\simeq10^8$\,yr. 
It is worth emphasizing that the momentum ratio $\zeta$ (eq.~\ref{momratio}) would then 
be overestimated.
The time-dependence of
a quasar might produce analogous effects; a fast outflow could be seen on 
large scales around a (now) less luminous AGN and a correspondingly low
value of $\zeta$ would be inferred.

{\bf Large $\tau_{\rm IR}$:} Using 2D planar gray flux-limited diffusion (FLD) with a 
realistic dust opacity Krumholz \& Thompson (2013) showed that 
the asymptotic momentum of shells driven with initially very large $\tau_{\rm IR}$
is not proportional to $\tau_{\rm IR}$.  In particular, for an initial midplane 
optical depth $\tau_{\rm IR}\simeq100-1000$ and IR Eddington ratio of $\infty$ 
(gravitational acceleration of zero),  they find that the asymptotic 
momentum taken up by the ejected material 
is only $\simeq1-10$ times that expected from the single-scattering
limit.  This result follows from the strong density-flux anti-correlation 
that develops as a result of channels which open in the 2D flow due in part to 
the radiation-driven Rayleigh-Taylor instability.

Recent results from \cite{davis_kt} using a more
sophisticated and accurate radiation transport algorithm (the Variable Eddington Tensor [VET] method;
\citealt{davis_short, jiang_rad}) supersede these earlier calculations, and produce
different results than FLD (see Krumholz \& Thompson 2012).
In particular, \cite{davis_kt}  find a less extreme flux-density anticorrelation that produces
more net momentum coupling between the radiation and the dusty gas relative to FLD.  This 
leads to a qualitatively different outcome in some simulations: Krumholz \& Thompson (2012)
find steady radiation pressure-driven convection whereas \citep{davis_kt} find an unbound
outflow for the same initial conditions.  

These results are important and should be more fully studied.   
There is yet no systematic study of the momentum coupling in super-Eddington 
dusty outflows with a large range of initial IR optical depths using multi-dimensional VET 
caclulations.\footnote{The related issue of the acceleration of individual clouds with dust opacity and accurate radiation 
hydrodyanmics remains to be investigated
(see \citealt{proga_cloud}).}  Such a study will be crucial in understanding the viability of radiation pressure in generating 
outflows with $\zeta\gg1$ in a range of contexts.  As implied by Figure \ref{figure:v} and the 
discussion in Section \ref{section:massive},
outbursts from massive stars will in general have large initial $\tau_{\rm IR}$ if dust forms.
In addition, we expect large average IR optical depths for massive star clusters and 
ULIRGs, and in the dusty pc-scale environments around AGN.  If radiation pressure is a
viable mechanism for the dynamics of these outflows then large effective IR optical depths 
for momentum coupling may be required by the data in some systems (e.g., Fig.~\ref{figure:c}).
These results from observational and numerical works may ultimately 
point either to other sources of wind driving, such as  
energy-driven flows powered by supernovae (e.g., \citealt{CC85,strickland_stevens,strickland_heckman}; but, see \citealt{zhang14}), 
or additional momentum input by supernovae (e.g., \citealt{mqt,tqm,faucher-giguere_feedback}), 
cosmic-rays (e.g., \citealt{jubelgas,socrates_cr,hanasz}), 
magneto-centrifugal acceleration, or other processes.

{\bf Grain Physics :} We have simplified a number of issues associated with grain physics.  First, we have neglected the temperature dependence of the Rosseland-mean opacity of dust grains in the optically-thick limit, relevant when $\tau_{\rm IR}>1$ \citep{pollack,semenov}.  In particular, $\kappa_R(T)\simeq2.4(T/{\rm 100\,K})^2$\,cm$^2/{\rm g}$ for $T\lesssim150$\,K and $\kappa_R(T)\sim{\rm const}$ for $150\lesssim T\lesssim1500$\,K for Milky Way gas-to-dust ratio.  Second, we have assumed that the gas and dust grains are completely dynamically coupled.  In reality, the momentum coupling between dust and gas is grain size dependent, and will depend both on the charge distribution and magnetic field strengths in the medium being accelerated \citep{draine_salpeter_2}.  Third, in the models presented throughout this paper, we assume that the grains in the gas are not destroyed by either the hard radiation fields of a central AGN or starburst, by grain-gas collisions, or sputtering (see, e.g., \citealt{draine_salpeter1}; \citealt{draine_salpeter_2,voit_grain,draine81,draine95}).  All of these issues deserve further investigation in the context of dusty radiation pressure accelerated shells.

\section{Summary}

We have shown that the typical expectation for radiation pressure driven flows
that $v_\infty \sim v_{\rm esc}$ at the launch radius $R_0$ is not correct for dusty 
shells and expanding clouds.  In these cases, the single-scattering phase of acceleration can dominate the 
dynamics.  For the shell case, the asymptotic velocity scales as 
$v_\infty\sim (R_{\rm UV} L/M_{\rm sh} c)^{1/2}
\propto\kappa_{\rm UV}^{1/4}(L/M_{\rm sh})^{1/2}$.  As discussed in Section \ref{section:introduction}
and \ref{section:dynamics} this result is equivalent to  $v_\infty \sim v_{\rm esc}\Gamma^{1/2}$ evaluated
at the point were the shell becomes optically-thin to the UV, $R_{\rm UV}$.   For clouds, the
dynamics is more complicated.  For individual rapidly expanding clouds (see Fig.~\ref{figure:clouds}), high 
velocities can be readily achieved, either through a long-lived single-scattering phase of acceleration, or 
because they become optically-thin to the UV radiation very rapidly so that they see a highly
super-Eddington flux.
The basic result $v_\infty$ can be very large has 
implications for the dynamics of dusty shells in a number of contexts,
including giant molecular cloud disruption around forming star clusters,
outbursts from massive stars, galactic winds
driven by star formation, and fast dusty outflows driven by AGN.  
In particular, it appears possible to accommodate the 
surprising result from \cite{diamond} that post-starburst galaxies 
with velocity dispersions of order $\sim200-250$\,km s$^{-1}$ can drive
shell-like outflows with velocity of $\sim2000$\,km s$^{-1}$ even though the
Eddington ratio at the launch radius was only $\Gamma_{\rm tot} \sim1$.

\section*{Acknowledgments}
The authors thank the anonymous referee for a timely and thoughtful report.
TAT is supported in part by NASA Grant NNX10AD01G.  TAT thanks
Aleks Diamond-Stanic, Nathan Smith, David Weinberg, Smita Mathur, 
and Dong Zhang for useful conversations and Chris Kochanek for a 
critical reading of the text.   EQ is supported in part by NASA ATP Grant 12-ATP12-0183.


\end{document}

%% file: journal.tex
\def\aj{AJ}%
\def\araa{ARA\&A}%
\def\apj{ApJ}%
\def\apjl{ApJ}%
\def\apjs{ApJS}%
\def\ao{Appl.~Opt.}%
\def\apss{Ap\&SS}%
\def\aap{A\&A}%
\def\mnras{MNRAS}%
\def\nat{Nature}%
